\documentclass[a4paper,fleqn,usenatbib]{mnras}

\usepackage{graphicx}
\usepackage{amsmath}
\usepackage{amssymb}
\usepackage{multicol}
\usepackage{bm}
\usepackage{pdflscape}
\usepackage[T1]{fontenc}
\usepackage{ae,aecompl}
\usepackage{mathptmx}
\usepackage[utf8]{inputenc}
\usepackage{algorithm}
\usepackage{algorithmic}

\newcommand{\cov}{\operatorname{cov}}
\newcommand{\diag}{\operatorname{diag}}
\newcommand{\dif}{\operatorname{d}}
\newcommand{\argmax}{\operatorname{argmax}}

\newcommand{\expect}{\operatorname{E}}
\newcommand{\gp}{\operatorname{GP}}

\pdfoutput=1

\title[GPE for dSphs]{Dynamical modelling of dwarf-spheroidal galaxies using Gaussian-process emulation}
\author[Gration \& Wilkinson]{
Amery Gration$^{1}$\thanks{E-mail: alg26@le.ac.uk} and Mark I.\ Wilkinson$^{1}$
\\
$^{1}$Department of Physics \& Astronomy, University of Leicester, University Road, Leicester LE1 7RH
}
\date{Accepted XXX. Received YYY; in original form ZZZ}
\pubyear{2018}

\begin{document}
\label{firstpage}
\pagerange{\pageref{firstpage}--\pageref{lastpage}}
\maketitle

\begin{abstract}
We present a novel and efficient method for fitting dynamical models of
stellar kinematic data in dwarf spheroidal galaxies (dSph). Our approach is
based on Gaussian-process emulation (GPE), which is a sophisticated form of
curve fitting that requires fewer training data than alternative methods. We
use a set of validation tests and diagnostic criteria to assess the
performance of the emulation procedure. We have implemented an algorithm in
which both the GPE procedure and its validation are fully automated. Applying
this method to synthetic data, with fewer than 100 model evaluations we are
able to recover a robust confidence region for the three-dimensional parameter
vector of a toy model of the phase-space distribution function of a
dSph. Although the dynamical model presented in this paper is low-dimensional
and static, we emphasize that the algorithm is applicable to any scheme that
involves the evaluation of computationally expensive models. It therefore has
the potential to render tractable previously intractable problems, for
example, the modelling of individual dSphs using high-dimensional,
time-dependent $N$-body simulations.
\end{abstract}


\section{INTRODUCTION}

In the $\Lambda$CDM model of cosmology, galaxies form by \emph{hierachical
  growth}, with large galaxies being formed by the agglomeration of smaller
ones. Of particular interest are the dwarf-spheroidal (dSph) galaxies that
orbit the Milky Way, the dark-matter haloes of which are the smallest dark
matter-dominated structures observed, and which are likely to be relics of the
earliest stages of galaxy formation. These galaxies are still not fully
understood. Although dark matter-only $N$-body simulations of halo-formation
using $\Lambda$CDM cosmology predict cusps in the dark-matter density profile
of the dSphs (i.e.\ they predict that density is inversely proportional to
radius for small radii), the literature on the modelling of observed dSphs
contains claims of both cusps and cores (i.e.\ it contains claims that density
may be constant for small radii) (\citealp{battaglia_kinematic_2008},
\citealp{strigari_kinematics_2010}, \citealp{breddels_model_2013}, and
\citealp{read_case_2018}). Either the $\Lambda$CDM model fails at small scales
and requires modification (\citealp{ludlow_2016,lovell_2014}), or dSphs evolve
over their lifetimes, under the influence of baryonic feedback. For example,
it has been shown that supernova-driven flattening may turn dark-matter cusps
into cores (\citealp{navarro_cores_1996}, \citealp{read_mass_2005}, and
\citealp{mashchenko_stellar_2008}). In the latter case we wish better to
understand the evolution of dSphs and as such require robust dynamical
modelling of their end-states to act as targets for evolutionary simulations.

To date, dynamical modelling of dSphs has for the most part been based on very
restrictive simplifying assumptions, namely that the dSphs are spherical and
in equilibrium (see, for example, \citealp{wilkinson_dark_2002},
\citealp{walker_universal_2009}, \citealp{strigari_kinematics_2010}, and
\citealp{read_how_2017}). While some authors have considered more general
models \citep{breddels_model_2013}, relaxation of these assumptions results in
models that are significantly more computationally expensive, often
prohibitively so. However, it is possible to reduce this computational expense
by borrowing the technique of \emph{emulation} from machine learning.

Let us establish some terminology. A \emph{simulation} is a model of some
phenomenon. Formally, a model is an indexed set of functions
$\{f(\cdot\,;a)\}_{a \in A}$ where $f(\cdot\,;a): X \longrightarrow Y$ is a
function, $a$ is called a \emph{parameter} and the set $A$ is called the
\emph{parameter space}. Specifically, in this paper, we assume some model of
the phase-space distribution function, $f(\cdot\,;a)$, of the stars within a
dSph, where $a$ parameterizes the physical model we are using. In this case,
$X$ is the phase space and $Y$ is the set of non-negative, real numbers.
A computer simulation is just a means of evaluating such a model, a simulation \emph{run} being an evaluation for a single parameter. In galactic dynamics we are typically interested in simulating (i.e.\ modelling) observations of a galaxy. 


When a model is computationally expensive to evaluate we may use a
\emph{meta-model}, i.e.\ a model of the model that is computationally
cheaper. An \emph{emulator} is a meta-model together with a measure of
confidence in that meta-model's output. In emulation we perform a small number
of runs, each using a different parameter, and then use the resulting data to
estimate the output for a parameter that we have not explicitly computed,
along with a confidence interval for that estimate. We do this without needing
to make an additional simulation run or model evaluation. In this respect,
emulation is a matter of \emph{prediction}. But, since we may make a
prediction for all points in the parameter space, we can fully map out the
function being emulated and in this respect emulation is an efficient means of
\emph{curve-fitting}.

One commonly used method of emulation is \emph{Gaussian-process emulation}
(GPE). In the astrophysical literature it has been used to fit exoplanetary
transit and secondary-eclipse light curves (\citealp{gibson_gaussian_2012} and
\citealp{evans_uniform_2015}), to map interstellar extinction within the Milky
Way \citep{sale_three-dimensional_2014}, and to fit semi-analytic models of
galaxy formation \citep{bower_parameter_2010}, while in the cosmological
literature it has has been used to predict the nonlinear matter power spectrum
in the Coyote Universe simulation \citep{heitmann_coyote_2009}, and to fit
Gravitational-wave models \citep{moore_improving_2016}. In the dynamical
modelling of stellar systems, GPE may allow us the use of more-expensive
static models or even full $N$-body models. In this paper we take a step
towards this goal by introducing the method and demonstrating its use with a
toy model, namely a single-component anisotropic Plummer sphere, which we fit
to synthetic data drawn from the same model. We focus on distribution
function-based models of the internal dynamics of a galaxy in order to
illustrate the value of the GPE approach. However, we note that in principle
the GPE approach is applicable to any dynamical modelling scheme.

To avoid confusion, a word of clarification is in order with respect to
terminology. The term `Gaussian' in `Gaussian-process emulation' refers only
to assumptions within the emulation scheme. It does not imply any assumptions
regards the Gaussianity, or otherwise, of any properties of the models
themselves. In particular, there is no assumption that the line-of-sight
velocity distributions are Gaussian, as is sometimes assumed in the dynamical
modelling of dSphs (\citealp{battaglia_kinematic_2008},
\citealp{strigari_common_2008}, and \citealp{ wolf_accurate_2010}).

Our principal interest is in what observed data can tell us about the
distribution of dark matter in a dSph. Which dark-matter morphologies do the
data rule out? Which best account for the observations? We therefore adopt the
maximum-likelihood approach, which allows us to draw robust confidence regions
within parameter space. We outline this procedure in
section~\ref{sec:likelihood}. In section~\ref{sec:gpe} we discuss the use of
GPE to estimate the likelihood when it too expensive to for us to perform a
parameter sweep. In section~\ref{sec:plummer} we illustrate the method using
our toy model. In section~\ref{sec:conclusion} we summarize our conclusions.

\section{Likelihood}
\label{sec:likelihood}

In this paper, our representation of a stellar system (in this case a dSph)
consists of a parameterized set of functions (i.e.\ the dynamical model)
representing the phase-space probability density function (also called the
distribution function). We treat the positions, $\bm{X}$, and the velocities,
$\bm{V}$, as random vectors, meaning that the state of a star is represented
by the random vector, $\bm{W} := (\bm{X}, \bm{V})$. The distribution function
(DF) for a single star is denoted $f_{\bm{W}}(\bm{w})$ \citep[for an extensive
  discussion, see][]{binney_galactic_2008}.\footnote{We adopt the notational
  convention that a random variable $\Omega$ (always capitalized) has PDF
  $f_\Omega$. Given a realization, $\omega$ (always lower case), of $\Omega$,
  and a parameter of the model, $a \in A$, this PDF then takes the value
  $f_{\Omega}(\omega; a)$.} We assume that the DF is an element of the model
$\{f(\bm{w};\bm{a})\}_{\bm{a} \in \bm{A}}$, where the parameters are
$D$-dimensional real vectors, the elements of which represent the total
galactic mass, galactic scale length, velocity anisotropy, etc. From the DF we
may calculate the observable properties of the system. For a dSph these
observables are typically the projected stellar positions, (represented by the
random variables $X$ and $Y$) and the line-of-sight velocity (represented by
the random variable $V_z$). The DF for these observables is given by the
marginalization of the phase-space DF:
\begin{align}
    f_{\bm{W}_{\mathrm{p}}}(\bm{w}_{\mathrm{p}}; \bm{a}) = \int_{\mathbf{R}^3}
    f_{\bm{W}}(\bm{w}; \bm{a}) \dif z \dif v_{x} \dif v_{y}.
\end{align}
We can reasonably assume that the states of stars are independent and
identically distributed \citep{binney_galactic_2008} so that the joint
marginalized PDF for $N$ stars is
\begin{align}
    f_{(\bm{W}_{\mathrm{p}, 1}, \ldots, \bm{W}_{\mathrm{p},
        N})}(\bm{w}_{\mathrm{p}, 1}, \ldots, \bm{w}_{\mathrm{p}, N}; \bm{a}) =
    \prod_{i = 1}^{N} f_{\bm{W}_{\mathrm{p}}}(\bm{w}_{\mathrm{p}, i}; \bm{a}).
\end{align}
By comparing these observables with a data set, we can then optimize the
parameters of the phase-space DF using the maximum likelihood method.

By definition, the likelihood of model parameter $\bm{a}$ is
\begin{align}
    L(\bm{a}; \bm{w}_{\mathrm{p}, 1}, \ldots, \bm{w}_{\mathrm{p}, N}) =
    f_{(\bm{W}_{\mathrm{p}, 1}, \ldots, \bm{W}_{\mathrm{p},
        N})}(\bm{w}_{\mathrm{p}, 1}, \ldots, \bm{w}_{\mathrm{p}, N}; \bm{a}).
    \label{eq:likelihood}
\end{align}
We recover the parameter by maximizing this function for given data, namely
the observed values of $\bm{w}_{\mathrm{p}, 1}, \ldots, \bm{w}_{\mathrm{p},
  N}$. We denote the maximum-likelihood estimate (MLE) by $\hat{\bm{a}}$. The
MLE is itself the realization of a random variable, which we denote
$\hat{\bm{A}}$. It is asymptotically normal (i.e.\ it is normally distributed
in the infinite-data limit) with mean $\bm{a}_0$ (the true parameter) and
variance $\bm{I}(\bm{a}_0)^{- 1}$, where the \emph{expected Fisher information
  matrix} is
\begin{align}
    \bm{I}(\bm{a}_0) = - \expect \left( \dfrac{\partial^2 \ln L(\bm{a}_0;
      \bm{W}_{\mathrm{p}, 1}, \ldots, \bm{W}_{\mathrm{p}, N})}{\partial \bm{a}
      \partial \bm{a}^{\mathrm{t}}} \right)
\end{align}
i.e.\ the negative of the expectation of the Hessian of the log-likelihood
\citep[see any text-book on the subject, for
  example][]{wasserman_all_2007}. The true parameter $\bm{a}_0$ may be
approximated by $\hat{\bm{a}}$, and the expected Fisher information matrix by
the \emph{observed Fisher information matrix},
\begin{align}
    \bm{J}(\hat{\bm{a}}) = - \dfrac{\partial^2 \ln L(\hat{\bm{a}};
      \bm{w}_{\mathrm{p}, 1}, \ldots, \bm{w}_{\mathrm{p}, N})}{\partial \bm{a}
      \partial \bm{a}^{\mathrm{t}}}.
    \label{eq:fisher_information}
\end{align}
In summary, the MLE is distributed as 
\begin{align}
    \hat{\bm{A}} \sim N(\hat{\bm{a}}, \bm{J}^{-1}(\hat{\bm{a}})).
\end{align}
This allows us to determine the confidence region, $C$, defined by the
boundary that is the solution to the equation
\begin{equation}
    (\bm{a} - \hat{\bm{a}})^\mathrm{t} \bm{J}^{-1}(\hat{\bm{a}}) (\bm{a} -
  \hat{\bm{a}}) = \chi_D^2(1 - \alpha)
    \label{eq:likelihood_confidence}
\end{equation}
where $\chi_D^2$ is the quantile function (i.e.\ the inverse of the cumulative
distribution function) for the chi-squared distribution for $D$ degrees of
freedom and $\alpha$ is the critical value, i.e.\ the value such that $C$
traps $\bm{a}_0$ with probability $\alpha$. The Fisher information quantifies
the curvature of the likelihood function at its maximum value and hence the
breadth of the distribution's peak. A narrow peak (i.e.\ large curvature and
large Fisher information) indicates that the maximum is well constrained. A
broad peak (i.e.\ small curvature and small Fisher information) indicates that
the maximum is poorly constrained.

Once we have the MLE of the parameters and the observed Fisher information
matrix, we may compute the distribution of the MLE of any function of the
parameters using the delta method \citep{wasserman_all_2007}. This states that
any real-valued function of the parameters, $g: \bm{A} \longrightarrow
\bm{R}$, is distributed normally with mean $g(\hat{\bm{a}})$ and variance
\begin{align}
    \dfrac{\partial g(\hat{\bm{a}})}{\partial \bm{a}^\mathrm{t}}
    \bm{J}^{-1}(\hat{\bm{a}}) \dfrac{\partial g(\hat{\bm{a}})}{\partial
      \bm{a}}.
    \label{eq:delta_method}
\end{align}
For us $g$ will be some function of the phase-space DF (for example the
galactic dark-matter profile, or Binney's anisotropy parameter).

The likelihood (equation~\ref{eq:likelihood}) will not usually be expressible
in closed form. Both the likelihood and its derivative may be costly to
evaluate. In such circumstances we may reduce the computational burden by
constructing a cheaper approximation of the likelihood using GPE.

\section{Gaussian-process emulation}
\label{sec:gpe}

At the heart of GPE \citep{ohagan_curve_1978, sacks_design_1989} are random
processes, which are sets of random variables. A realization of a random
process (i.e. a realization of each individual random variable) is a function,
meaning that we may use Gaussian-processes in the context of
curve-fitting. More rigorously, a random process, $\{Y(\bm{a})\}_{\bm{a} \in
  \bm{A}}$, is an indexed set of random variables. Again, we call an index,
$\bm{a}$, a \emph{parameter} (or \emph{parameter vector} if $\bm{A}$ is
multidimensional), and we call the index set, $\bm{A}$, the \emph{parameter
  space}. A random process is \emph{Gaussian} if every finite subset of the
process has a multivariate normal distribution. A process is completely
specified by the joint PDF for every finite subset of the process. Since the
joint PDF of a multivariate normal distribution is completely specified by its
mean and covariance, a Gaussian process is completely specified by its mean
and covariance. The mean is the function $m: A \longrightarrow \bm{R}$ such
that $m(\bm{a}) = \expect(Y(\bm{a}))$. The covariance is the function $k:
\bm{A}^2 \longrightarrow \bm{R}$ such that $k(\bm{a}, \bm{a}') = \cov(\bm{a},
\bm{a}')$. If a process, $Z$, is Gaussian we write
\begin{align}
  Z \sim \mathrm{GP}(m(\bm{a}), k(\bm{a}, \bm{a}')).
\end{align}

Recall that in curve-fitting (or, \emph{regression} as it is called in the
statistical literature) we have a random variable $\bm{A}$ (called the
independent variable) and a random variable $Y$ (called the dependent
variable), and wish to find the relationship between them. Given
\emph{training data} $((\bm{a}_{i}, y(\bm{a}_{i})))_{i = 1}^N$ we may always
write the relationship in the form
\begin{align}
    Y(\bm{a}_i) = r(\bm{a}_i) + E(\bm{a}_i)
    \label{eq:regression_model}
\end{align}
where the \emph{regression function} is $r(\bm{a}) := \expect(Y(\bm{a}))$ for
all $\bm{a}$ \citep{wasserman_all_2007}, and $E(\bm{a}_i)$ is a random
variable (the \emph{noise}.

We seek an estimator for $r$, denoted $\hat{r}$. In \emph{parametric}
regression we assume that this estimator is an element of some parameterized
set of functions, $\hat{r} \in \{f_{\bm{\beta}} \,|\, \bm{\beta} \in
\bm{B}\}$. We call this set the \emph{regression model} and the set $\bm{B}$
the set of \emph{regression parameters}. For example, in linear regression we
assume that this model is $\{\beta_{0} + \sum_{i = 1}^{D} \beta_{i} a_{i}
\,|\, \beta_{0}, \ldots, \beta_{D} \in \bm{R}\}$. Regression is
\emph{non-parametric} if we may not write the estimator in this
way. Non-parametric regression is useful when we have no motivation for a
parameterized expression for the regression function. In Gaussian-process
regression we assume that
\begin{align}
    Y \sim \gp(r, \sigma^2)
    \label{eq:gpe_assumption}
\end{align}
whereupon $r = \expect(Y)$, as required. Then $E(\bm{a}) = Y(\bm{a}) -
r(\bm{a})$, i.e.\ $E$ is a Gaussian process with zero mean. We will assume
that its covariance is the sum of two terms, i.e.\ that $\sigma^{2}(\bm{a},
\bm{a}') = k(\bm{a}, \bm{a}') + \sigma^{2}_{\epsilon}(\bm{a})$, the first term
representing the signal and the second term representing the noise (i.e.\ the
error in the dependent variable). Note that the noise is not in general
constant, but is a function of $\bm{a}$. In statistical parlance, we would say
that the errors are \emph{heteroscedastic} rather than \emph{homoscedastic}.

Suppose we wish to predict the dependent variable $Y(\bm{a})$ for some value
$\bm{a} \in \bm{A}$. By hypothesis, the distribution of this random variable
is univariate normal, i.e.\
\begin{align}
    Y(\bm{a}) \sim N(r(\bm{a}), \sigma^{2}(\bm{a}, \bm{a})),
\end{align}
and the joint distribution of the sample is multivariate normal, i.e.
\begin{align}
    \bm{Y} \sim N(\bm{r}, \bm{K}),
    \label{eq:sample_distrib}
\end{align}
where $\bm{r} := (r(\bm{a}_{1}), r(\bm{a}_{2}), \ldots, r(\bm{a}_{N}))$, and $[\bm{K}]_{ij} = k(\bm{a}_{i}, \bm{a}_{j}) + \sigma_{\epsilon}^{2}(\bm{a}_{i})\delta_{ij}$.
The estimator for our regression function is the mean of the conditional random variable
\begin{align}
    Y(\bm{a}) \,|\, (\bm{Y} = \bm{y}) \sim N(\hat{r}, \hat{\sigma}^2)
    \label{eq:gpe_conditional}
\end{align}
where $[\bm{y}]_{i} = y(\bm{a}_{i})$, i.e.\ the expected value of $Y(\bm{a})$ given the training data. It is a standard result \citep{ohagan_curve_1978} that
\begin{align}
  \hat{r}(\bm{a})
  &= r(\bm{a}) + \bm{k}^\mathrm{t}(\bm{a})\bm{K}^{- 1} (\bm{y} - \bm{r}), \text{ and}
  \label{eq:gpe_mean} \\
  \hat{\sigma}^2(\bm{a})
  &= k(\bm{a}, \bm{a}) - \bm{k}^\mathrm{t}(\bm{a})\bm{K}^{- 1} \bm{k}(\bm{a}) \label{eq:gpe_var}
\end{align}
where $[\bm{k}(\bm{a})]_i = k(\bm{a}_i, \bm{a})$. These two equations are the principal results of GPE. The estimator, $\hat{r}$, is the sum of two terms. The first is the regression function, and the second a smoothing of the residuals, $(\bm{y} - \bm{r})$. In fact this smoothing is a weighted sum of the residuals, where the weights are the elements of the vector $\bm{k}^\mathrm{t}(\bm{a})\bm{K}^{- 1}$.

\subsection{Optimizing the emulator}
\label{sec:optimizing_hyperparams}

To evaluate equations~\ref{eq:gpe_mean} and~\ref{eq:gpe_var}, we must assume a mean function, $r$, and a covariance function, $k$. We will choose these from a model, i.e.\ we will assume that $r \in \{r_{\bm{\mu}} \,|\, \bm{\mu} \in \bm{M}\}$ and $k \in \{k_{\bm{\nu}} \,|\, \bm{\nu} \in \bm{N}\}$, where we call $\bm{M}$ and $\bm{N}$ sets of \emph{hyperparameters}. We optimize our choice of hyperparameters using their maximum likelihood. By equation~\ref{eq:sample_distrib} the PDF is
\begin{align}
    \begin{split}
    &f_{\bm{Y}}(\bm{y}; \bm{\theta})\\
    &\quad = \dfrac{1}{\sqrt{(2 \pi)^{N} |\bm{K}|}} \exp \left(-\dfrac{1}{2} (\bm{y} - \bm{r})^{\mathrm{t}} \bm{K}^{-1} (\bm{y} - \bm{r}) \right)
    \end{split}
\end{align}
where the hyperparameter vector $\bm{\theta} := (\bm{\mu}, \bm{\nu})$. (Note that $\bm{K}$ and $\bm{r}$ depend on $\bm{\theta}$.) By definition, the likelihood is
\begin{align}
    L_{\bm{\Theta}}(\bm{\theta}; \bm{y}) = f_{\bm{Y}_{N}}(\bm{y}; \bm{\theta})
\end{align}
and hence
\begin{align}
    &\ln L_{\bm{\Theta}}(\bm{\theta}; \bm{y}) = - \dfrac{1}{2} (\bm{y} - \bm{r})^{\mathrm{t}} \bm{K}^{-1} (\bm{y} - \bm{r}) - \dfrac{1}{2} \ln |\bm{K}| - \dfrac{N}{2} \ln 2 \pi
    \label{eq:hyperparam_max_likelihood}
\end{align}
\citep{rasmussen_gaussian_2006}. To find the maximum-likelihood estimate of $\bm{\theta}$, namely $\hat{\bm{\theta}} = (\hat{\bm{\mu}}, \hat{\bm{\nu}})$, we may maximize this function subject to the constraint that $\bm{K}$ is positive-semidefinite (positive-semidefiniteness being a necessary property of covariance matrices). The equation for $\ln L_{\Theta}(\bm{\theta}; \bm{y})$ consists of three terms. The first is a measure of fit quality, the second is a complexity penalty, and the third a normalization constant. 

The complexity penalty is a function of $\bm{K}$ only, and quantifies the complexity of our attempted fit independent of the data. For a complicated fit, the covariance of any two points is low. Hence the determinant of the covariance matrix $\bm{K}$ is small, and $\ln |\bm{K}|$ diverges with complexity (i.e.\ as $|\bm{K}| \longrightarrow 0$ so $\ln |\bm{K}| \longrightarrow - \infty$). This strongly penalizes complex models. \cite[For a full discussion see][section 5.4.1.]{rasmussen_gaussian_2006} The function $L_{\Theta}$ will in general have multiple maxima, each maximum giving a different tradeoff between fit quality and complexity. 

The function $r_{\bm{\mu}}$ might be a finite linear combination of basis functions, $(\xi_i)_{i = 1}^N$, and $\bm{\mu}$ the set of coefficients:
\begin{align}
    r(\bm{a}; \bm{\mu}) 
    &= \sum_{i = 1}^{N} \mu_{i} \xi_{i}(\bm{a})\\
    &= \bm{\mu}^{\mathrm{t}} \bm{\xi}(\bm{a})
    \label{eq:gpe_basis}
\end{align}
where $[\bm{\mu}]_{i} = \mu_{i}$ and $[\bm{\xi}(\bm{a})]_{i} = \xi_{i}(\bm{a})$. However, we are free to choose any function for $r$. Specifically we may choose the zero function, such that $r(\bm{a}) = 0$ for all $\bm{a} \in \bm{A}$, whereupon the Gaussian-process fit to the residuals (the second term in equation~\ref{eq:gpe_mean}) does all the work of the regression. It is common practice to do this \cite[][section 2.7]{rasmussen_gaussian_2006}, and we will do so for the rest of this paper. \footnote{We note that in \cite{bower_parameter_2010}  $r$ is not taken to be the zero function, but instead is a sum of polynomials in the elements of $\bm{a}$.}


 
The covariance function, $k_{\bm{\nu}}$, is usually chosen from one of a number of standard functions. It is a function of the parameter space $\bm{A}$ and we will need to impose some structure of this space. For our purposes, $\bm{A} = \bm{R}^D$, treated as linear vector space of dimension $D$, together with a pseudo-metric, $d: \bm{A}^2 \longrightarrow \bm{R}$, i.e.\ a symmetric positive-semidefinite function of two variables that obeys the triangle inequality. \footnote{Parameter space is dimensional meaning that a naive definition of a metric on the space $\bm{R}^D$ has no meaning. However, we may impose a metric on the space of functions defined by the phase-space PDF, which then induces a pseudo-metric on the parameter space used to index these functions. A pseudo-metric is similar to a metric, but the condition of positive-definiteness is relaxed to become positive-semidefiniteness, i.e.\ two nonidentical elements of the space may be separated by zero distance.} It is common to assume that the pseudo-metric may be written in quadratic form, i.e.\ it is common to assume that 
\begin{align}
    d^2(\bm{a} - \bm{a}') = (\bm{a} - \bm{a}')^\mathrm{t} \bm{M} (\bm{a} - \bm{a}'),
    \label{eq:metric}
\end{align}
for positive-semidefinite matrix $\bm{M}$.

The most common covariance function is the \emph{squared-exponential} (SE),
\begin{equation}
    k_{\mathrm{SE}}(\bm{a}, \bm{a}')     = \sigma_{\mathrm{SE}}^2 \exp \left(-\dfrac{1}{2} d^2(\bm{a} - \bm{a}') \right),
    \label{eq:cov_SE}
\end{equation}
where $\sigma_{\mathrm{SE}}^{2}$ is the \emph{signal variance} (the value of $k_\mathrm{SE}$ when $\bm{a}' = \bm{a}$). Furthermore, it is normally assumed that the pseudo-metric matrix is diagonal, i.e.\ that $\bm{M} = \diag(m_{1}, m_{2}, \ldots , m_{D})$. Using equations~\ref{eq:metric} and ~\ref{eq:cov_SE} we may show \citep[see, for example,][]{loeppky_choosing_2009} that the mean squared gradient is 
\begin{equation}
    \expect \left( \frac{\upartial Y(\bm{a})}{\upartial a_i} \right)^2 = 2 \sigma_\mathrm{SE}^2 m_i, 
\end{equation}
and therefore call $m_i$ the \emph{sensitivity} of the model to the $i$-th parameter. Because $\bm{M}$ is positive-semidefinite it has a unique positive-semidefinite inverse, which in turn has a unique square root, i.e.\ there exists a unique matrix $\bm{L}$ (not to be confused with the likelihood, $L$) such that $\bm{M} = \bm{L}^{-2}$. If $\bm{M}$ is diagonal then so is $\bm{L}$ and $\bm{L} = \diag(l_1, l_2, \ldots , l_D)$ where $m_i = l_i^{-2}$ for all $i$. We can see that equation~\ref{eq:cov_SE} is formally identical to a Gaussian with covariance $\bm{M}^{-1}$. We therefore identify $\bm{L}^{2}$ as a covariance matrix and call the element $l_i$ the \emph{correlation length} (also \emph{scale length}) for the $i$-th parameter. The elements of $\bm{M}$ together with $\sigma_{\mathrm{SE}}^2$ and the coefficients of our basis functions form the hyperparameter vector, $\bm{\theta} = \{\sigma_{\mathrm{SE}}^2, \mu_{1}, \ldots, \mu_{N}, m_{1}, \ldots , m_{D}\}$, of the Gaussian-process estimator. One advantage of assuming that $r$ is the zero function is that this hyperparameter vector becomes shorter:  $\bm{\theta} = \{\sigma_{\mathrm{SE}}^2, m_{1}, \ldots , m_{D}\}$. This means that the dimension of the domain of the hyperparameter likelihood is smaller and hence the hyperparameter vector is easier to optimize. 

If $r$ is expanded according to equation~\ref{eq:gpe_basis} and we use the squared-exponential covariance function then the mean-square error of the estimator, i.e.\ the mean-square difference of $\hat{r}$ and $r$ (Sacks et al.\ 1989), is 
\begin{align}
    \begin{split}
    &\operatorname{MSE}(Y(\bm{a}) \,|\, (\bm{Y} = \bm{y})) = \\
    &\sigma_\mathrm{SE}^2
    \bigg(1 -
    \begin{bmatrix}
        \bm{\xi}^{\mathrm{t}}(\bm{a}) & \sigma_\mathrm{SE}^{-2} \bm{k}^{\mathrm{t}}(\bm{a})
    \end{bmatrix}
    \begin{bmatrix}
        \bm{0}      & \bm{\Xi}^{\mathrm{t}}\\
        \bm{\Xi}   & \sigma_\mathrm{SE}^{-2} \bm{K}
    \end{bmatrix}^{\mathrm{-1}}
    \begin{bmatrix}
        \bm{\xi}(\bm{a})\\
        \sigma_\mathrm{SE}^{-2} \bm{k}(\bm{a})
    \end{bmatrix}
    \bigg)
    \end{split}
    \label{eq:mse}
\end{align}
where $[\bm{\Xi}]_{i} = \bm{\xi}^{\mathrm{t}}(\bm{a})$. In the case that $\bm{\xi}(\bm{a}) = \bm{0}$ (as we assume here) this expression reduces to the estimator variance (equation~\ref{eq:gpe_var}). We may therefore construct a confidence interval for the estimator, $\hat{r}(\bm{a})$, using the root mean-square error (RMSE), namely the interval
\begin{align}
  C = \left(\hat{r}(\bm{a}) - \Phi^{-1}(1 - \alpha / 2) \hat{\sigma}(\bm{a}), \hat{r}(\bm{a}) + \Phi^{-1}(1 - \alpha / 2) \hat{\sigma}(\bm{a})\right)
  \label{eq:confidence_interval}
\end{align}
where $\Phi$ is the CDF for the univariate normal distribution and $\alpha$ is the critical value. If there are no errors associated with the training data then the RMSE of the estimator is zero at the training points and increases as the distance of the test point from a training point increases. If the errors on the sample are constant, then we expect the RMSE to be approximately constant (except at the boundaries where it will be greater on account of boundary bias).

\subsection{Validating the emulator}
\label{sec:validating_the_emulator}

The machinery of Gaussian-process regression assumes that the covariance function, $k$, is known. In practice, it never is. We must chose an approximation to the covariance, invariably from a list of standard covariance functions, as discussed above. We would like to know that this covariance function has been well chosen, i.e.\ we would like to assess the performance of the emulator given our choice of covariance function. We may do this using leave-one-out cross-validation (LOOCV) \citep{wasserman_all_2007}. We note that the estimator (equation~\ref{eq:gpe_mean}) is distributed normally with variance equal to the MSE (equation~\ref{eq:mse}), and expect that for a well-specified covariance this distribution will be observed in our estimator. Ideally we would evaluate our dynamical model at a large number of test points, and compare the distribution of the residuals of our predictions with that of the MSE. This, of course, imposes an impractical computational burden. Instead, we omit the $i$-th pair, $(\bm{a}_i, y(\bm{a}_i))$, from our training data to give the reduced data, $(\bm{a}_i, y(\bm{a}_i))_{i \neq i}$. Using these data we then compute an estimate for $y$ at the omitted point, $\bm{a}_i$, finding that (see equation~\ref{eq:gpe_conditional}) $Y(\bm{a}_i) | (\bm{Y}_{-i} = \bm{y}_{-i}) \sim N(\hat{y}_{-i}(\bm{a}_i), \hat{\sigma}_{-i}^2(\bm{a}_i))$. This gives us $N$ residuals, which we may compare with the MSE. We expect that the \emph{standardized predicted error} \cite{wasserman_all_2007},
\begin{equation}
    e_{-i}(\bm{a}_i) := \frac{y(\bm{a}_i) - \hat{r}_{-i}(\bm{a}_i)}{\hat{\sigma}_{-i}(\bm{a}_i)},
    \label{eq:predicted_error}
\end{equation}
is normally distributed with zero mean and unit variance.\footnote{In practice, we do not need to calculate the LOOCV residuals directly. It can be shown (\citealp{sundararajan_predictive_2001}) that the standardized predicted error, $e_{-i}(\bm{a}_i) = [\bm{K}^{-1} \bm{y}]_{ii} / \sqrt{[\bm{K}]_{ii}}$, which regrettably requires the explicit inversion of $\bm{K}$.}

One way to compare two distributions is by means of a \emph{quantile-quantile plot}, in which we plot the quantiles of one distribution against the quantiles of the other. If the two distributions are the same, then the points will lie on the diagonal. If either distribution is empirical we may substitute its ordered observed values for its quantiles. To check that the standardized residuals are distributed as required, we therefore plot the $N$ ordered residuals of our fit against the $N$-th quantiles of the normal distribution. If the covariance function has been well-specified we expect them to be evenly scattered along the diagonal. We also require that the residuals are small, and hence calculate the \emph{mean squared predicted error} (also \emph{leave-one-out cross-validation score}),
\begin{equation}
    R := \frac{1}{N} \sum_{i = 1}^{N} (y(\bm{a}_i) - \hat{y}_{-i}(\bm{a}_i))^2,
    \label{eq:cv_score}
\end{equation}
which tends to zero in the infinite-training data limit, as well as plot the predicted values against the true values. Furthermore, we require that the residuals exhibit no trend, and hence plot the standardized predicted errors against the associated predicted values for $y$. Finally, we require that there are no significant outliers (greater than three sigma, or $e_{-i} > 3$, say), which would indicate that the estimator is underperforming is certain regions of parameter space.

We expect poor accuracy in our estimator when neighbouring points are poorly correlated, i.e.\ when the scale of features in our function is approximately equal to or less than the point separation. We also expect the accuracy to be poor at the boundary of parameter space, where the model is constrained by data on one side only. We may consider a point to be near the boundary of parameter space if it is within one parameter length scale of it.

We may in fact use LOOCV to optimize the hyperparameter vector, as opposed to the maximum-likelihood method described in section~\ref{sec:gpe}. Note that the logarithm of the likelihood of the hyperparameter vector when leaving out the $i$-th pair of data (called the \emph{leave-one-out likelihood}) is
\begin{align}
    \ln L_{-i, \bm{\Theta}}(\bm{\theta}; \bm{y}) = - \dfrac{1}{2} \ln \sigma_{-i}^2(\bm{a}_i) - \dfrac{1}{2} e_{-i}^2(\bm{a}_i) - \dfrac{1}{2} \ln 2 \pi
\end{align}
\citep{rasmussen_gaussian_2006}. Hence the logarithm of the joint leave-one-out likelihood (LOO likelihood) is
\begin{align}
    \ln L_{\mathrm{LOO}, \bm{\Theta}} = \sum_{i = 1}^N \ln L_{\mathrm{pseudo}, \bm{\Theta}}(\bm{\theta}; \bm{y}),
\end{align}
which we may maximize to give the \emph{maximum LOO likelihood estimate} of the hyperparameter vector. This estimate is more robust to mispecification of the covariance function \citep{bachoc_cross_2013} than the maximum-likelihood estimate (equation~\ref{eq:hyperparam_max_likelihood}).

\subsection{Conditioning the likelihood}
\label{sec:conditioning}
If the estimator fails validation then the covariance function has been misspecified. In this case we may do one of two things: choose a different covariance function, or transform the data so that the function is better suited to its task. Here we restrict ourselves to using the squared-exponential covariance function (equation~\ref{eq:cov_SE}), and hence consider the second approach. 

We have assumed that our data is drawn from a Gaussian random process with covariance $k_{\mathrm{SE}}(\bm{a}, \bm{a}') + \sigma_\epsilon^{2}(\bm{a})$ (equation~\ref{eq:gpe_assumption}). Crucially, the squared exponential covariance function is a function of the difference of its arguments only and the properties of the random process are therefore independent of the absolute values its arguments. In particular, the variance and autocovariance of the random process are constant. The fact that the variance is constant means that noise-free data are identically normal. If we do not observe this distribution in our training data we may transform it to ensure that we do. Such a transformation is said to be \emph{variance stabilizing}.

One such variance-stabilizing transformation is the \emph{Box-Cox transformation}. Let $\{y(\bm{a}_i)\}_{i = 1}^N$ be a set of observations. Then the Box-Cox transformation of the observations \citep{box_analysis_1964} is the function $g$ such that
\begin{align}
  g(y(\bm{a}_i); \lambda_1, \lambda_2) = \dfrac{(y(\bm{a}_i) + \lambda_2)^{\lambda_1} -
    1}{\lambda_1}
\end{align}
for some real $\lambda_1$, $\lambda_2$ such that $\lambda_{2} > - y(\bm{a}_{i})$ for all $i$. Note that this expression is just a scaled power law, with the scaling chosen such that $\lim_{\lambda_1 \longrightarrow 0} g(y(\bm{a}_i)) = \ln(y(\bm{a}_i) + \lambda_2)$.

Box and Cox give the following theorem concerning the choice of the parameters $\lambda_1$ and $\lambda_2$. Assume that each observation $y(\bm{a}_{i})$ is a realization of the random variable $G_i$, and that each transformed observation, $g(y(\bm{a}_{i}))$ is a realization of the random variable $G'_{i}$. Furthermore assume that $G'_{1}, \ldots, G'_{N}$ are independent and identically normal, i.e.\ assume that for all $n$,
\begin{align}
  G'_{i} \sim N(\mu_\mathrm{t}^{\phantom{2}}, \sigma_\mathrm{t}^2).
  \label{Box_Cox_assumption}
\end{align}
for some mean, $\mu_\mathrm{t}^{\phantom{2}}$, and variance $\sigma_\mathrm{t}^2$. Then the joint distribution of the (untransformed) observations has PDF
\begin{align}
    \begin{split}
    &f_{(G_{1}, \ldots, G_{N})}(y(\bm{a}_{1}), \ldots, y(\bm{a}_{N}); \lambda_1, \lambda_2) \\
    &= \prod_{i = 1}^{N} \dfrac{1}{\sqrt{2 \pi \sigma_\mathrm{t}^2}} \exp \left( \dfrac{\sum_{i = 1}^{N} (g(y(\bm{a}_{i})) - \mu_\mathrm{t})^2}{2 \sigma_\mathrm{t}^2} \right) J(y(\bm{a}_{i}); \lambda_1, \lambda_2),
    \end{split}
\end{align}
where the Jacobian
\begin{align}
    J(y(\bm{a}_{i}); \lambda_1, \lambda_2)
    &= \left| \dfrac{\dif g(y(\bm{a}_i))}{\dif y(\bm{a}_i)} \right|\\
    &= (y(\bm{a}_i) + \lambda_2)^{(\lambda_1 - 1)}.
\end{align}
Thus the log-likelihood of the parameters $\lambda_1$ and $\lambda_2$ is given by
\begin{align}
    \begin{split}
        &\ln L_{(\lambda_1, \lambda_2)}(\lambda_1, \lambda_2; y(\bm{a}_{1}), \ldots, y(\bm{a}_i)) = - \frac{N}{2} \ln \sigma_\mathrm{t}^2 - \frac{N}{2}\log 2 \pi \\
        & \quad - \frac{1}{2\sigma_\mathrm{t}^2}\sum_{i = 1}^{N} (g(y(\bm{a}_i)) - \mu_\mathrm{t})^2 + (\lambda_1 - 1) \sum_{i = 1}^{N} \ln(y(\bm{a}_i) + \lambda_2)
    \end{split}
    \label{eq:Box_Cox_likelihood}
\end{align}
where we may substitute for $\mu_\mathrm{t}^{\phantom{2}}$ and $\sigma_\mathrm{t}^2$ their maximum-likelihood estimates,
\begin{align}
    \hat{\mu}_{\mathrm{t}} &= \frac{1}{N} \sum_{i = 1}^N g(y(\bm{a}_i)) \text{ and}\\
    \hat{\sigma}_\mathrm{t}^2 &= \dfrac{1}{N} \sum_{i = 1}^N (g(y(\bm{a}_i)) -
    \hat{\mu}_{\mathrm{t}})^2.
\end{align}
The MLE of the transformation parameters has an associ ated distribution, but it is common to consider it known, and not propagate this uncertainty through the subsequent analysis. It is also common to round off the value of $\lambda_1$ to the nearest half-integer, e.g.\ to use one of the values $2, 1, 1 / 2, 0, -1 / 2, -1,$ or $-2$ (the square, identity, square root, logarithm, reciprocal square root, reciprocal, or reciprocal square) which give the transformation a ready interpretation. The natural transformation of the likelihood (\ref{eq:likelihood}) is to its logarithm together with an offset that ensures this logarithm is defined when the likelihood is zero. In this case the likelihood of the offset, $\lambda_2$, is strictly decreasing, and its optimization will fail. In such a case we may use some arbitrary small value for $\lambda_2$, say the smallest non-zero member of the sample.

We must be careful about the direction of the implication in this theorem. It is not the case that the maximum-likelihood estimates of $\lambda_1$ and $\lambda_2$ ensure that the transformed variables are normal. We must compute these maximum-likelihood estimates and then \emph{check} that the assumption of normality is met by inspecting a histogram of the transformed values. In general the condition is not met, i.e.\ there exists no power transformation such that $G'_{1}, \ldots, G'_{N} \sim N(\mu_\mathrm{t}, \sigma_\mathrm{t}^2)$. However we may still use the power transformation to \emph{regularize} $G$, i.e.\ to bring its distribution closer to normal \citep{draper_distributions_1969}. 

When using the zero regression function the residuals $\bm{y} - \bm{r}$ (equation~\ref{eq:gpe_mean}) may be poorly behaved. They may all have the same sign, or may contain outliers. The use of a transformation overcomes these problems, and avoids the need to specify a regression function, of which we have very little prior knowledge.

Having stabilized the variance of the data, we must also ensure the
autocovariance is approximately constant. We do this by transforming the
parameter space (i.e.\ by reparameterizing the random process). This is an
altogether more difficult problem \citep[see, for
  example][]{sampson_nonparametric_1992} as the length scale $l_i$ (and hence
sensitivity $m_i$) is a function of the parameter $a_i$. We will assume that
for positive parameters the sensitivity $m_i$ is monotonically decreasing in
$a_i$ and converges on zero, i.e.\ we will assume that the model is sensitive
to changes in the parameters when they are small, but insensitive when they
are large, or, equivalently, that the likelihood is slowly varying for
arbitrarily large parameter values, when it is approximately zero, but quickly
varying when the parameter values are small, when it is significantly
non-zero. We therefore make a logarithmic transformation of the parameters. If
our assumption is wrong, this transformation will not improve the accuracy of
our predictions. We must therefore rely on the validation process to tell us
if we have made a useful transformation. We must accept that the covariance
function will always be misspecified to some degree. If the misspecification
is so significant that the model fails validation, we have to choose a
different covariance function, i.e.\ we have to choose a covariance function
that is not the squared-exponential. We do not consider such a situation here,
but note that there is a large literature on the subject \citep[see, for
  example][]{rasmussen_gaussian_2006}.

\subsection{Training the emulator}
\label{sec:training}

Assuming we are able to choose some region of parameter space for which we wish emulate the likelihood, we must chose a \emph{design} (i.e.\ an arrangement of points in parameter space at which we will sample the function). If \emph{a priori} we know nothing about our function (beyond our assumptions of continuity and smoothness encoded in our choice of covariance function) we wish the design to be \emph{space-filling}, i.e.\ to have uniform density throughout parameter space. We also wish all projections of the design onto lower-dimensional subspaces to be space-filling, as the model may have low sensitivity to some parameters. Lattices are a poor choice for such a design, as their size grows exponentially with the dimension of the parameter space. The most commonly-used designs satisfying the above space-filling requirements are \emph{Latin hypercubes} \citep{mckay_comparison_1979}. In Latin hypercube sampling (LHS), a parameter space is partitioned into a hypergrid of $N^D$ cells and $N$ points are placed in these cells such that there is only one point in any hyper-row or hyper-column of cells. We may optimize the space-filling property of the design by maximizing the minimum separation of pairs of points in all projections of the design onto lower-dimensional subspaces \citep{santner_design_2003}. LHS designs are restricted to rectangular regions. It is possible to form space-filling designs on nonrectangular regions (e.g.\ \citealp{draguljic_noncollapsing_2012}) but we do not consider these here. 

We choose the size of our design, $N$, so that the GPE model has acceptable accuracy. This size depends on the difficulty of the problem, i.e.\ the complexity of the function we are emulating. The more difficult the problem, the greater $N$ will need to be. The question obviously arises: how do we choose an appropriate value of $N$ for a particular problem? 

We note that $N$ is satisfactory if the MSE (equation~\ref{eq:mse}) is small, and that the MSE is a function of $\bm{M}$ (through the covariance function, $k$), $N$ (through the size of the matrix $\bm{K}$, and $D$ (the dimensional of the parameter space). We wish to understand the relationship between these quantities. To this end, Loeppky \emph{et al.}\ (2008) introduce the \emph{total sensitivity},
\begin{equation}
    \label{eq:tau}
    \tau := \sum_{i = 1}^D m_i,
\end{equation}
and the \emph{sparsity},
\begin{equation}
    \label{eq:psi}
    \psi := \sum_{i = 1}^D m_i^2,
\end{equation}
where $\{m_i\}_{i = 1}^D$ is the set of elements of the metric matrix. Recall that the length scale $l_i$ is defined such that $m_i = l_i^{-2}$. Consider the squared separation of a pair of design points, $\bm{a}_i$ and $\bm{a}_j$, namely $d^2(\bm{a}_i, \bm{a}_j)$. For a random LHS design, this separation is the realization of a random variable, $H$. Loeppky \emph{et al.}\ show that for such a design $H$ is distributed with mean $E(H) = \mu(N) \tau$ and variance $\operatorname{var}(H) = \nu(N) \psi$ where $\mu$ and $\nu$ are weak and strictly decreasing functions of $N$ that converge to a positive constant. The accuracy of our emulator will be good when $E(H)$ is small (i.e.\ when the mean separation of sample points is small, and hence mean correlation is good) and when $\operatorname{var}(H)$ is large (i.e.\ when many pairs of points have separations smaller than the mean and are hence even better correlated).

If we minimize $\psi$ whilst keeping $\tau$ constant (i.e.\ if we minimize $\psi$ subject to the constraint $\sum_i m_i = c$ for some real $c$) we find that $m_j = c / D$ for all $j$, i.e.\ we find that $\psi$ is a minimum (and hence the accuracy poor) when the parameters are equally active, and hence $\psi = c^2 / D$. On the other hand, if we maximize $\psi$ whilst keeping $\tau$ constant we find that $m_j = c$ for some $j$ and $m_i = 0$ for $i \neq j$, i.e.\ we find that $\psi$ is a maximum (and hence the accuracy good) when only one parameter is active, and hence $\psi = c$. For fixed $\tau$, therefore, $\psi$ quantifies the sparsity of the matrix $\bm{M}$. In the case that all parameters are equally active it is the case that $E(H) = D c$ and that $\operatorname{\psi} = D c^2$, i.e.\ that both the mean and the variance of the separation are proportional to the number of parameters, and that for a sufficiently large number, the accuracy will be poor. 

The accuracy of our estimator depends on both the total sensitivity and the sparsity. It does not depend on the total number of parameters but rather on the number of active parameters. Suppose that the parameter space has been mapped to a hypercube of side $h$. Motivated by practical experiment, Loeppky \emph{et al.}\ propose that if $\tau h^2 = 3$ then the problem is ``easy'', and if $\tau h^2= 40$ the problem is ``very difficult''. If $\tau h^2 = 10$ the the problem will be tractable if $\psi$ is small but intractable if $\psi$ is large. Moreover, for easy problems the convergence of $R$ to one is fast whereas for difficult problems the convergence is slow. As a rule of thumb, easy problems will have good accuracy for $N = 10D$, whereas difficult problems will require significantly greater $N$. 

Training is therefore best done iteratively. We first take a sample of size $10D$ and validate the emulator. If the model accuracy is poor the covariance function is misspecified, in which case we must use a different covariance function, or an unduly large amount of training data. Due to the slow convergence of the MSE in this case (i.e.\ the case where a sample size of $10D$ is too small), we will need to resample the function in a smaller region of parameter space. If the model accuracy is good, we augment our data. To do this we require some figure of merit for choosing new design points. If we wish to emulate the function faithfully across the region, we might resample at points of maximum variance. If we wish to maximize the function, as we do here, we may use the \emph{expected improvement}. In this case the procedure is known as \emph{efficient global optimization} \citep{jones_efficient_1998}.

\subsection{Improving the emulator}
\label{sec:improving_the_emulator}

We follow the presentation of \cite{schonlau_global_1996}, which we reproduce here in our own notation, for clarity. In the mathematical literature, optimization problems are couched in terms of minimization rather than maximization. We adopt this convention here, understanding of course that we may maximize a function by minimizing its negative. 

Suppose that we are performing Gaussian-process regression (equations~\ref{eq:regression_model}, \ref{eq:gpe_assumption}) using training data $(\bm{a}_i, y(\bm{a}_i))_{i = 1}^{N}$. The minimum of our response sample, $y_\mathrm{min} = \min(y(\bm{a}_i))_{i = 1}^{N}$. We would like to know where to sample in order to improve the accuracy of this minimum. To this end we define the \emph{improvement in the minimum},
\begin{align}
  I_Y(Y) := \max(y_\mathrm{min} - Y, 0).
\end{align}
This is a random variable, the PDF of which is
\begin{align}
  I_Y(y) &= \max(y_\mathrm{min} - y, 0)\\ &=
  \begin{cases}
    y_\mathrm{min} - y &\text{if $y < y_\mathrm{min}$,}\\ 0 &\text{otherwise}.
  \end{cases}
\end{align}
By definition the expected improvement is
\begin{align}
  \expect(I_Y(Y)) := \int_{\bm{R}} I_Y(y) f_Y(y) \dif y,
\end{align}
where $f_Y$ is the PDF of $Y$. In the case of GPE we know that $Y \sim N(\hat{y}, \hat{\sigma}^2)$, i.e.\ $f_Y$ is the normal (i.e.\ Gaussian) PDF $\phi(y; \hat{y}, \hat{\sigma}^2)$. If $\hat{\sigma}^2 = 0$ then the value $y$ is known with certainty and we cannot expect any improvement, hence $\expect(I_Y(Y)) = 0$. If $\hat{\sigma}^2 > 0$ then we may make the change of variables from $y$ to $u' = (y - \hat{y}) / \hat{\sigma}$ to find that the expected improvement is
\begin{align}
  \expect(I_Y(Y)) &= \hat{\sigma} \int_{-\infty}^{u}(u - u') \phi(u'; 0, 1) \dif u'
\end{align}
where $u := (y_\mathrm{min} - \hat{y}) / \hat{\sigma}$. Thus, 
\begin{align}
  \expect(I_Y(Y)) =
  \begin{cases}
    \hat{\sigma} (u \Phi(u; 0, 1) + \phi(u; 0, 1))
    \label{EI_expression} &\text{ if $0 < \hat{\sigma}$,}\\
    0 &\text{ otherwise}
  \end{cases}
\end{align}
where $\Phi(u; 0, 1)$ is the normal cumulative distribution function. We augment our training data, with the pair $(\bm{a}_{N + 1}, f(\bm{a}_{N + 1}))$ where $\bm{a}_{N + 1} = \argmax(\expect(I_Y(Y)))$, and then iterate this procedure until $\expect(I_Y(Y))$ is smaller than some threshold, $\epsilon$. Efficient global optimization (EGO) is implemented by Algorithm~\ref{algorithm}.

\begin{algorithm}[t]
  \caption{Efficient global optimization}
  \label{algorithm}
  \begin{algorithmic}
    \REQUIRE objective function, $f$, sample of objective function, $D := \{(\bm{x}_i, y(\bm{x}_i))\}_{i = 1}^N$, and stopping threshold, $\epsilon$
    \ENSURE global minimum of objective function
    \STATE $E_\mathrm{max} \leftarrow \max(\expect(I))$
    \WHILE{$E_\mathrm{max} > \epsilon$:}
    \STATE $D \leftarrow D \cup (\argmax(\expect(I)), f(\argmax(\expect(I)))$
    \STATE $E_\mathrm{max} \leftarrow \max(\expect(I))$
    \ENDWHILE
    \RETURN $\bm{x}_i$ such that $y(\bm{x}_i) = \min (y(\bm{x}_i))_{i = 1}^N$.
  \end{algorithmic}
\end{algorithm}

The expected improvement for $0 < \hat{\sigma}^2$ is the sum of two terms in $u$. The first term dominates if $u$ is large, while the second term dominates if $u$ is small. For given $\hat{y}$ it is the case that $u$ is large if $\hat{\sigma}$ is small (which will be the case close to design points, including the current minimum) and $u$ is small if $\hat{\sigma}$ is large (which will be the case away from design points, including the current minimum). The expected improvement is therefore a tradeoff between probable small improvements (near to the current minimum) and improbable large improvements (remote from the current minimum), or between local and global search. The fact that the expected improvement is a trade off between local and global search makes multistart optimization a sensible choice if we initialize it at each design point. We may use gradient-based methods as the gradient of the expected improvement has closed form. 

The efficient global optimization algorithm reduces the problem of the prohibitively-expensive optimization of $y$ to the cheap optimization of the expected improvement. There is a convergence theorem \citep{vazquez_convergence_2010} that guarantees that the expected improvement produces a sequence of points that is dense in the parameter space under mild assumptions about the covariance function, so that the result is guaranteed to be a global minimum in the infinite-sample limit. However, we do not know of any theorems concerning the rate of convergence.

\begin{figure}
    \centering
    \includegraphics[width=1.0\linewidth]{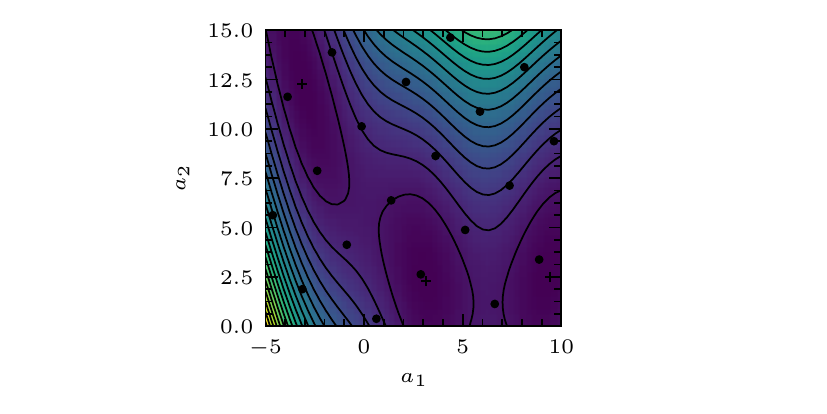}
    \caption{The Branin function (equation~\ref{eq:branin}) and the Latin
      square design (marked with filled circles) used in its
      emulation. Following \protect\cite{jones_efficient_1998}, we use it to
      illustrate the methods of Gaussian process emulation
      \protect\citep{ohagan_curve_1978} and efficient global optimization
      \protect\citep{jones_efficient_1998}. It has three global minima (marked with
      crosses).}
    \label{fig:branin_function}
\end{figure}

\begin{figure}
    \centering
    \includegraphics[width=1.0\linewidth]{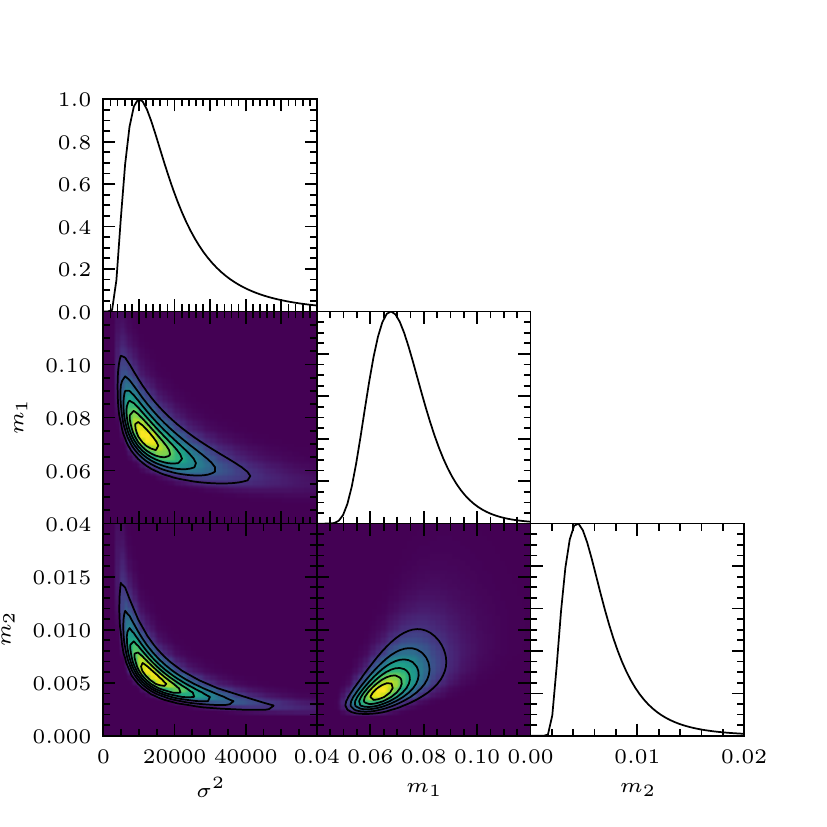}
    \caption{The likelihood (equation~\ref{eq:hyperparam_max_likelihood}) of the hyperparameters of the squared-exponential covariance function (equation~\ref{eq:cov_SE}), used in the emulation of the Branin function (equation~\ref{eq:branin}). In each panel the marginal likelihood is shown (i.e. the likelihood has been integrated over the parameters not shown), scaled to the unit interval. The maximum likelihood is found at $(\sigma_\mathrm{SE}^2, m_, m_2, m) = (15\thinspace500, 0.0689, 0.00520)$, i.e.\ for length scales, $l_1 = 3.81$ and $l_2 = 13.9$.}
    \label{fig:branin_hyperparameter_likelihood}
\end{figure}

\begin{figure}
    \centering
    \includegraphics[width=1.0\linewidth]{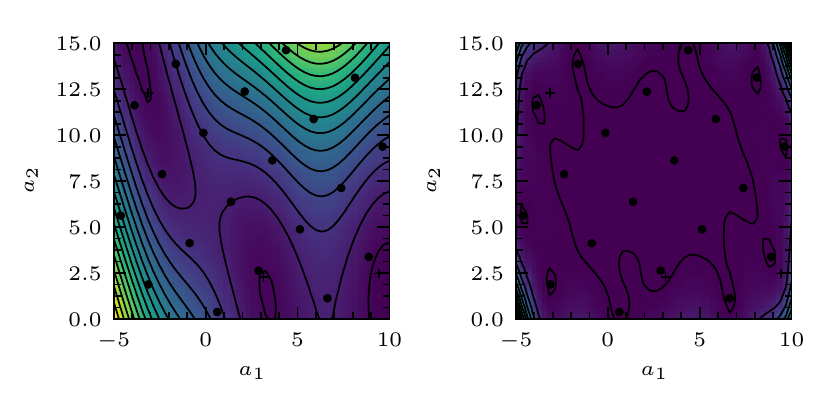}
    \caption{The mean (left) and variance (right) of GPE estimate of the Branin function, computed using the squared-exponential covariance function and the design shown (filled circles). The variance is high in regions of parameter space that have been poorly sampled, or near the boundary of parameter space, where the estimator is constrained by less data than elsewhere.}
    \label{fig:branin_function_gpe}
\end{figure}

\begin{figure}
    \includegraphics[width=1.0\linewidth]{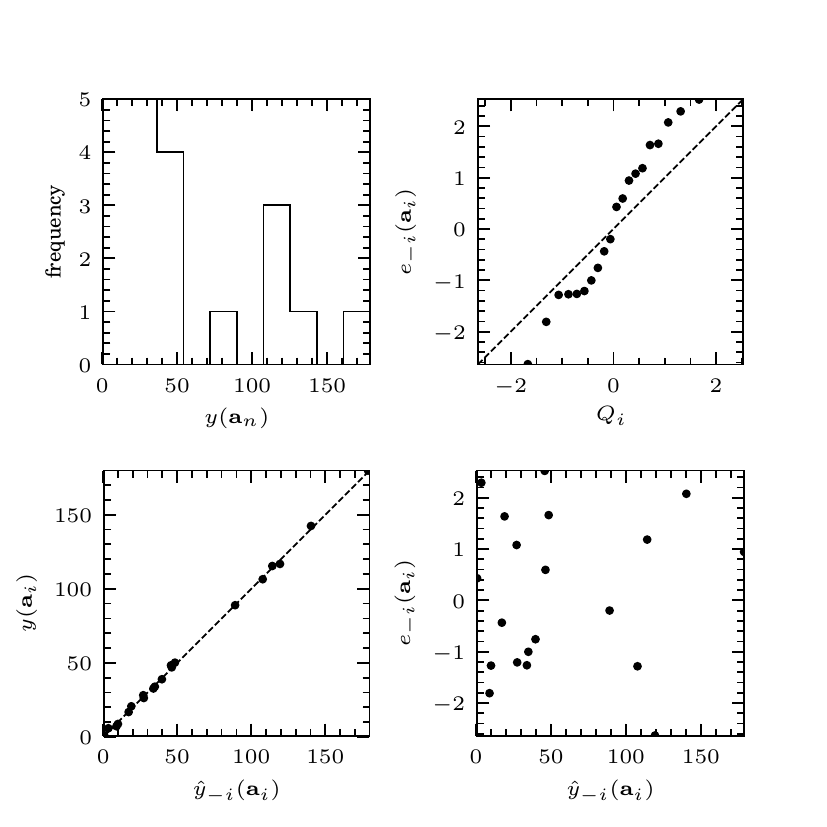}
    \caption{Diagnostic plots for the emulation of the Branin function. Top-left: distribution of likelihood values. Top-right: quantile-quantile plot showing the ordered standardized predicted errors from a LOOCV analysis (equation~\ref{eq:predicted_error}) against the equivalent quantiles of the normal distribution. Bottom-left: true values against predicted values from a LOOCV analysis. Bottom-right: residuals from LOOCV analysis.}
    \label{fig:branin_function_diagnostics}
\end{figure}

\begin{figure}
    \centering
    \includegraphics[width=1.0\linewidth]{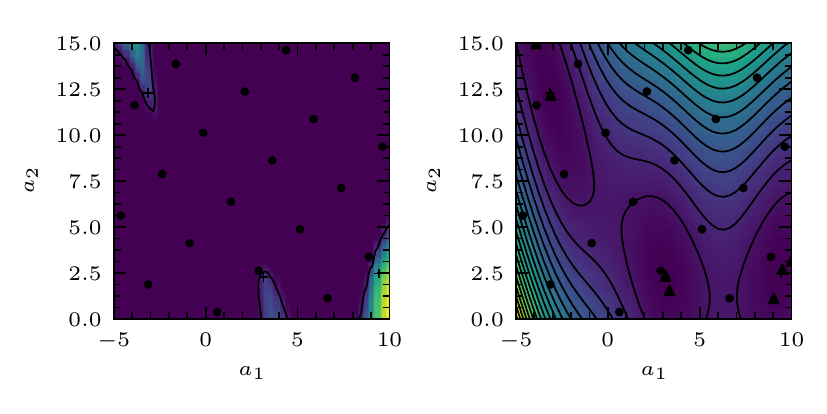}
    \caption{The expected improvement in the minimum  (left) of our sample of the Branin function, computed using the design shown (filled circles). There are three maxima, each close to a minimum of the Branin function. The Branin function and the augmented design (marked with filled triangles) determined by the EGO algorithm (right). The new data cluster about the function's three minima. Once the initial design has been computed only 8 additional design points are required to find a global minimum with an accuracy of $1.3$ per cent.}
    \label{fig:branin_function_augmented}
    \label{fig:branin_function_expected_improvement}
\end{figure}

We illustrate EGO by reproducing an example given by \citeauthor{jones_efficient_1998}, namely the minimization of the \emph{Branin function}, a real-valued function of two variables used as a test for optimization. For the sake of completeness, we also produce figures equivalent to theirs. The Branin function,
\begin{equation}
    y(a_1, a_2) = \alpha (a_2 - \beta^2 + \gamma a_1 - \delta)^2 + \zeta(1 - \eta) \cos a_1 + \eta,
    \label{eq:branin}
\end{equation}
where $\alpha = 1$, $\beta = 5.1 / (4 \pi^2)$, $\gamma = 5 / \pi$, $\delta =
6$, $\zeta = 10$, $\eta = 1 / (8 \pi)$. It has three global maxima, at $(a_1,
a_2) = (-\pi, 12.275)$, $(\pi, 2.275)$, and $(9.425, 2.475)$ where the
function takes the value $0.398$. It is evaluated on the domain $a_1 \in [-5,
  10]$, $a_2 \in [0, 15]$. We treat the function as a realization of a random
process, $\{Y_{\bm{a}}\}_{\bm{a} \in \bm{A}}$, where the parameter space,
$\bm{A} = [-5, 10] \times [0, 15]$. We create a LHS design for the parameter
space, namely the set $\{\bm{a}_i\}_{i = 1}^N$, of size $N = 10D = 20$ and
evaluate the function, $y$, at these points, giving the data $\{\bm{a}_i,
y(\bm{a}_i)\}_{i = 1}^N$. The function and the design are plotted in
Figure~\ref{fig:branin_function}. We assume a squared-exponential covariance
function, $k_\mathrm{SE}(\bm{a}, \bm{a}') = \sigma_{\mathrm{SE}}^2 \exp (-
(\bm{a} - \bm{a}')^\mathrm{t} \bm{M} (\bm{a} - \bm{a}') / 2)$ where $\bm{M} =
\diag(m_1, m_2)$ (see equation~\ref{eq:cov_SE}), and then optimize its
hyperparameters, $\sigma_\mathrm{SE}^2$, $m_1$, and $m_2$, using the maximum
likelihood method (equation~\ref{eq:hyperparam_max_likelihood}), finding that
$\sigma^2 = 15\thinspace500$, $m_1 = 0.0689$, and $m_2 = 0.00520$, or
equivalently that $l_1 = 3.81$, $l_2 = 13.9$. For the sake of illustration, we
plot the likelihood of the hyperparameters in
Figure~\ref{fig:branin_hyperparameter_likelihood} as well as the mean and
variance of the Gaussian-process estimator for the entire parameter space in
Figure~\ref{fig:branin_function_gpe} (see equations~\ref{eq:gpe_mean},
\ref{eq:gpe_var}). We also compute the standardized mean square predicted
error, finding that $R = 2.01$ (equation~\ref{eq:cv_score}), and the maximum
standardized predicted error, finding that the most extreme value of $e_{i}$
is 2.63computational (see equation~\ref{eq:predicted_error}). Diagnostic plots
are shown in Figure~\ref{fig:branin_function_diagnostics}. We see that the
standardized predicted errors are distributed normally and exhibit no trend
across the parameter space. The total sensitivity, $\tau h^2 = 16.7$, is
moderate, explaining this success. Satisfied that our model is accurate, we
then iteratively augment the data set using the maximum expected improvement
and a stopping criterion of $\epsilon = 0.001$. For illustration we plot the
expected improvement for the first iteration
(Figure~\ref{fig:branin_function_expected_improvement}) and see that it has
three maxima close to the three minima of the Branin function. The algorithm
requires a total of eight iterations to find the maximum to an accuracy of
$1.3$~per cent . We plot the augmented design in
Figure~\ref{fig:branin_function_augmented}.

As with any global optimization method, it is sensible to polish the result, which may lack accuracy. We may do so by resampling the function in the neighbourhood of this result, and again performing regression. The length scales of the GPE fit provides a guide to the size of this neighbourhood. If $\bm{x}_0 = (x_{0, 1}, x_{0, 2}, x_{0, 3})$ is the result of the EGO algorithm then we resample the function in the region $\bm{A}' = [x_{0, 1} - \delta l_1, x_{0, 1} + \delta l_1] \times [x_{0, 2} - \delta l_2, x_{0, 2} + \delta l_2] \times [x_{0, 3} - \delta l_3, x_{0, 3} + \delta l_3]$, where $(l_0, l_1, l_2)$ is the vector of length scales found in the final iteration, and $\delta$ is some positive real number less than one. We may again use GPE to perform this regression. Our polished maximum is then the maximum of the GPE estimator. We can find this maximum using gradient-based optimization, or, as the GPE estimator is cheap, by brute force i.e.\ by searching over a fine lattice of test points covering the whole region $\bm{A}'$. (This brute force method returns more the maximum, of course. It maps out the function over the entirety of $\bm{A}'$. In general the cheapness of the GPE estimator will allow us to do just this. When working in very high-dimensional parameter spaces the inversion the covariance matrix, $\bm{K}$, may not be so cheap, and we may wish to map out the region using, for example, Markov chain Monte Carlo methods.) There are two additional benefits to this polishing step. First, we may use a high termination threshold, $\epsilon$, which reduces the number of iterations required by the EGO algorithm. Second, the Hessian of the GPE estimate is available in closed form, and provides an estimate of the Hessian of the function. If the function in question is a likelihood, this allows us to compute an estimate of the Fisher information matrix. The derivative of a Gaussian process is itself a Gaussian process \citep{adler_geometry_2010}:
\begin{align}
    \dfrac{\partial Y}{\partial \bm{a}} \sim \gp \left( \dfrac{\partial \hat{r}}{\partial \bm{a}}, \dfrac{\partial^2 k(\bm{a}, \bm{a}')}{\partial \bm{a} \partial \bm{a}^\mathrm{\prime t}} \right).
    \label{eq:GPE_deriv}
\end{align}
If $H = \partial^2 y / \partial \bm{a} \partial \bm{a}^\mathrm{t}$ is the Hessian of the function $y$, then an estimate for the Hessian is 
\begin{align}
    \hat{H} = \dfrac{\partial^2 \hat{r}}{\partial \bm{a} \partial \bm{a}^\mathrm{t}}.
    \label{eq:gpecompute_hessian}
\end{align}
We compute the Hessian for the case of the squared-exponential covariance function in Appendix~\ref{sec:se_derives}.

\subsection{Computational expense}

The computational complexity of equations~\ref{eq:gpe_mean} and \ref{eq:gpe_var} is dominated by the inversion of the matrix $\bm{K}$, which is of order $O(N^3)$ or better. This inversion may be done indirectly by solving the systems $\bm{K} \bm{\alpha} = \bm{y}$ and $\bm{K} \bm{\beta} = \bm{k}(\bm{a})$ for $\bm{\alpha}$ and $\bm{\beta}$ respectively. Moreover, it must be performed only once, regardless of the number of test evaluations required. Once the inversion has been performed, evaluation of the mean involves one matrix-vector multiplication, with complexity $O(N^2)$, followed by one vector-vector multiplication for each evaluation, with complexity $O(N)$. Each evaluation of the variance involves one matrix-vector multiplication and one vector-vector multiplication. Thus the computational complexity is $O(N^3)$. The covariance matrix $\bm{K}$ must also be inverted for every step in the optimization of the hyperparameters, which must be done at each iteration of the EGO algorithm. Furthermore, it must be computed explicitly for the validation step. Nonetheless, the total computational expense of these inversions is negligible compared with the expense of any interesting astrophysical simulation. The expense of the method is largely in the computation of the training data, and its augmentation required when using the EGO method. The initial sampling is trivially parallel, and linear in the number of parameters, but the EGO method is necessarily sequential. (A batch-sequential extension of EGO is available, and makes it possible to perform up to 10 function evaluations at each iteration.) In general we cannot estimate in advance the number of iterations required for EGO without knowing the rate of convergence of the EGO algorithm, so we do not explore this further here.

\section{A TOY APPLICATION}
\label{sec:plummer}

\subsection{The anisotropic Plummer sphere}

\begin{figure}
    \centering
    \includegraphics[width=1.0\linewidth]{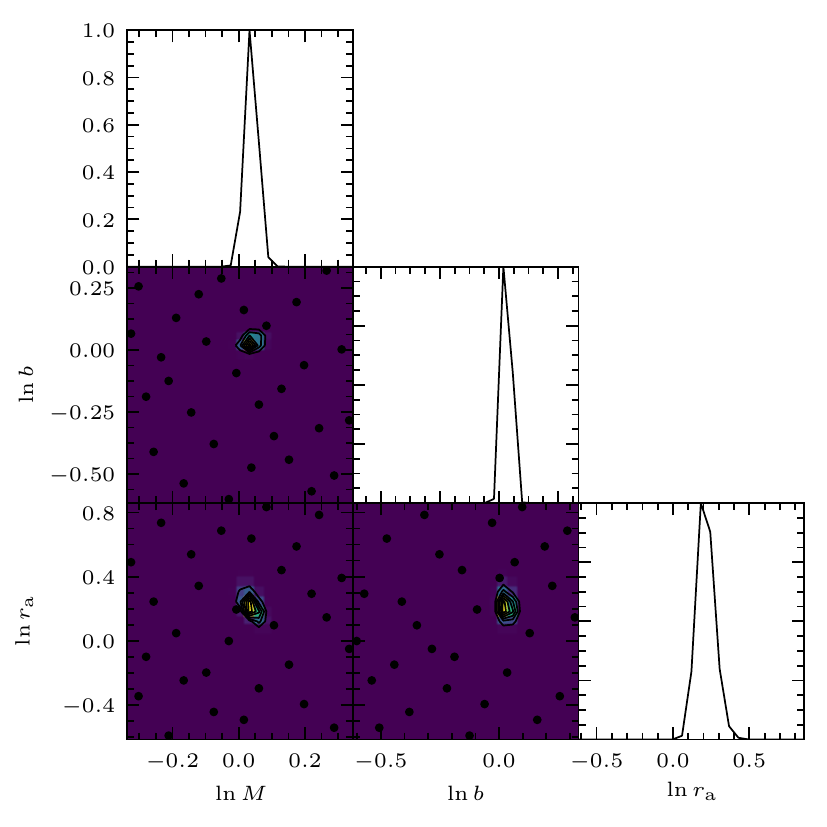}
    \caption{The marginalized likelihood, $L$, for the anistropic Plummer model with parameters $\log M = 0$, $\log b = 0$, $\log r_\mathrm{a} = \log 2 = 0.301$ computed using data for 1000 stars generated by the same model (equation~\ref{eq:OM_line_profile}). In each panel the likelihood has been marginalized over the unshown parameters, and scaled to the unit interval. The maximum likelihood is found at $\log M = 0.0327$, $\log b = -0.0213$, $\log r_\mathrm{a} = \log 2 = 0.305$. We do not emulate this function directly, but rather its Box-Cox transformation $\ln(L + \lambda)$, where $\lambda$ is an arbitrary small constant, here taken to be the smallest non-zero element of our sample of $L$. The design used in the emulation of $L$ is shown with filled circles.}
    \label{fig:plummer_likelihood}
    \centering
    \includegraphics[width=1.0\linewidth]{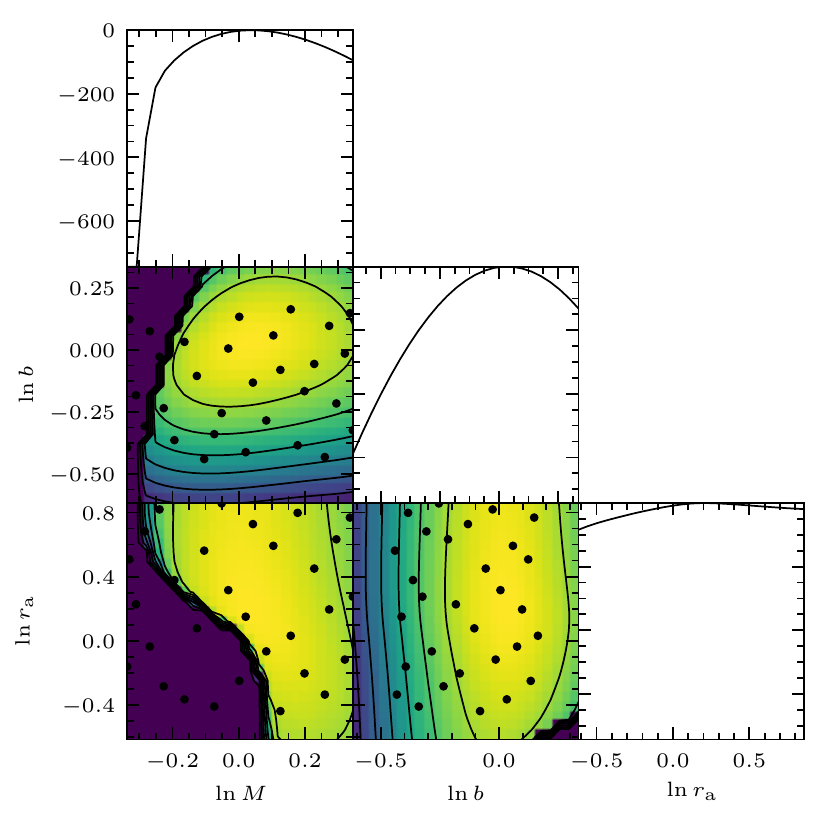}
    \caption{The log-marginalized likelihood, $L$, for the anistropic Plummer model with parameters $\log M = 0$, $\log b = 0$, $\log r_\mathrm{a} = \log 2 = 0.301$ computed using data for 1000 stars generated by the same model (equation~\ref{eq:OM_line_profile}). In each panel the likelihood has been marginalized over the unshown parameters, scaled to the unit interval, and its natural logarithm plotted.}
    \label{fig:plummer_log_likelihood}
\end{figure}


We illustrate the use of GPE for stellar dynamical modelling using the toy model of an anisotropic Plummer sphere of the Osipkov-Merritt type. The relative potential of the Plummer sphere,
\begin{align}
  \Psi(r) = \dfrac{G M}{\sqrt{r^2 + b^2}},
\end{align}
and its density
\begin{align}
  \rho(r) = \dfrac{3 M}{4 \pi b^3}\left( 1 + \frac{r^2}{b^2} \right)^{- 5 / 2}
  \label{eq:density}
\end{align}
where $M$ is the galactic mass, $r$ the radius, and $b$ the galactic scale length \citep{plummer_problem_1911}.

Following \cite{osipkov_spherical_1979} and \cite{merritt_spherical_1985}, we define the variable $Q = \mathcal{E} - L^2 / 2 r_\mathrm{a}^2$ where $\mathcal{E}$ is the relative energy, $L$ is the magnitude of the angular momentum, and $r_\mathrm{a}$ is the anisotropy radius. By use of Eddington's inversion formula we then find that the phase-space DF may be expressed as a function of $Q$:
\begin{align}
  f_Q(Q) = \dfrac{3 M b^2}{\pi^3 \sqrt{2} r_\mathrm{a}^2} \left( \dfrac{16(r_\mathrm{a}^2 - b^2)}{7}
    Q^{7 / 2} + (G M)^2 Q^{3 / 2}\right).
  \label{eq:OM_pdf}
\end{align}

The PDF of the observables is given by the integral of $f_Q(Q)$ with respect to the line-of-sight position and proper-motion velocities. If we define the parameter vector $\bm{a} = (M, b, r_\mathrm{a})$ and work in cylindrical coordinates with the $z$-axis parallel to the line of sight, this PDF
\begin{align}
  f_{(R_{\mathrm{p}}, V_z)}(r_{\mathrm{p}}, v_z; \bm{a}) = 2 \pi \int_{\bm{R}}
  \int_{\bm{R}} \int_{\bm{R}} f_Q(Q) \dif v_{r_\mathrm{p}} \dif
  v_{\phi} \dif z.
  \label{eq:OM_line_profile}
\end{align}
We seek to maximize the likelihood $L(\bm{a}; r_\mathrm{p}, v_z) := f_{(R_{\mathrm{p}}, V_z)}(r_\mathrm{p}, v_z; \bm{a})$. The inner double integral may be computed analytically using the method given by \cite{carollo_velocity_1995}:
\begin{align}
\int_{\bm{R}}
  \int_{\bm{R}} f_{Q}(Q) \dif v_{r_{\mathrm{p}}} \dif v_\phi = 
\begin{cases}
    2 \pi g(r, r_{\mathrm{p}}) F(Q_{\mathrm{max}}) &\text{ if $0 < Q_{\mathrm{max}}$,}\\ 
    0 &\text{ otherwise}
\end{cases}
\end{align}
where
\begin{align}
  g(r_{r, \mathrm{p}}) &:= \dfrac{a^2}{\sqrt{(r_\mathrm{a}^2 + r^2)(r_\mathrm{a}^2 + r^2 - r_\mathrm{p}^2)}},\\
  F(Q) &:= \dfrac{6 b^2}{\pi^3 \sqrt{2} r_\mathrm{a}^2 (G M)^5} \left(\dfrac{16(r_\mathrm{a}^2 - b^2)}{63} Q^{9 / 2} + \dfrac{(GM)^2}{5} Q^{5 /
      2}\right),
\end{align}
and
\begin{align}
  Q_{\mathrm{max}}(r_\mathrm{p}, z, v_{z}) = \Psi(r) - \left(\dfrac{r_\mathrm{a}^2 + r^2}{r_\mathrm{a}^2 + r^2 - r_\mathrm{p}^2}\right)
  \dfrac{v_{z}^2}{2}.
\end{align}
However, the outer integral in equation~\ref{eq:OM_line_profile} must be computed numerically.

\subsection{Optimization of the likelihood}

\begin{figure}
    \centering
    \includegraphics[width=1.0\linewidth]{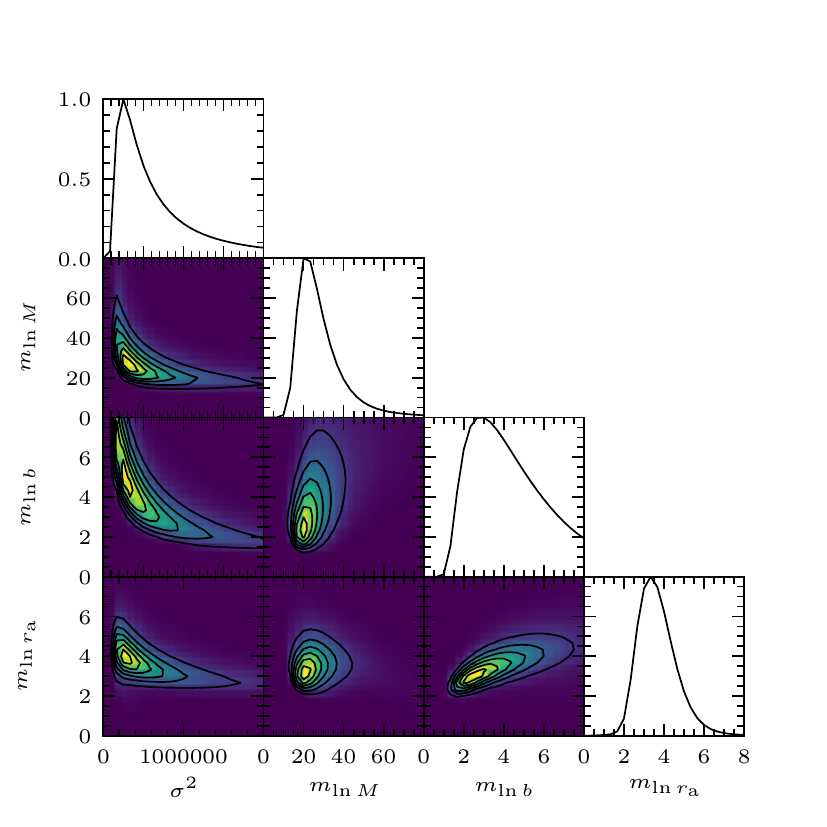}
    \caption{The likelihood (equation~\ref{eq:hyperparam_max_likelihood}) of the hyperparameters of the squared-exponential covariance function (equation~\ref{eq:cov_SE}), used in the emulation of the Plummer-model likelihood (equation~\ref{eq:OM_line_profile}). In each panel the marginal likelihood is shown (i.e. the likelihood has been integrated over the parameters not shown), scaled to the unit interval. The maximum likelihood is found at $(\sigma_\mathrm{SE}^2, m_{\log M}, m_{\log b}, m_{\log{r_\mathrm{a}}}) = (535\thinspace000, 23.5, 3.30, 3.43)$, i.e.\ for length scales, $l_{\log M} = 0.206$, $l_{\log b} = 0.550$, and $l_{\log{r_\mathrm{a}}} = 0.540$).}
    \label{fig:hyperparameter_likelihood}
\end{figure}

\begin{figure}
    \includegraphics[width=1.0\linewidth]{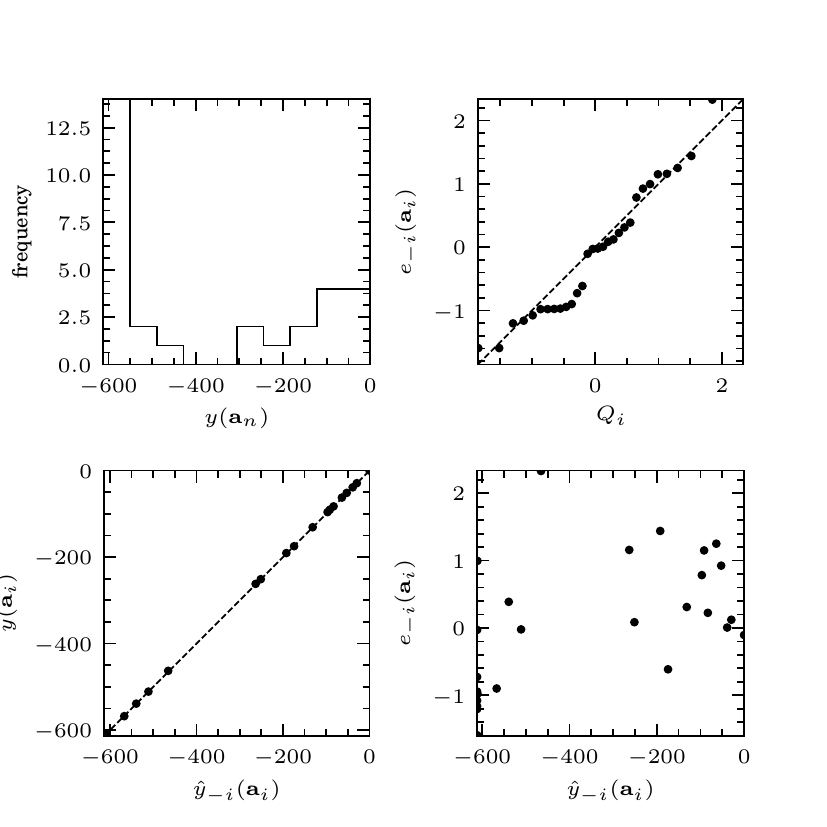}
    \caption{Diagnostic plots for the emulation of the transformed Plummer model likelihood. Top-left: distribution of likelihood values. Top-right: quantile-quantile plot showing the ordered standardized predicted errors from a LOOCV analysis (equation~\ref{eq:predicted_error}) against the equivalent quantiles of the normal distribution. Bottom-left: true values against predicted values from a LOOCV analysis. Bottom-right: residuals from LOOCV analysis.}
    \label{fig:plummer_function_diagnostics}
\end{figure}

We work in mass units of $10^9 \text{M}_{\sun}$, and distance units of kpc (meaning that the gravitational constant $G = 4.302 \times 10^{3} \text{kpc} \text{M}_{\sun}^{- 1} \text{km}^{2} \text{s}^{- 2}$). We use synthetic data generated using the same model and parameters $M = 1$, $b = 1$, and $r_\mathrm{a} = 2$. The data consist of positions and line-of-sight velocities for 1000 stars, each with zero error. In the case of the anistropic Plummer model the likelihood is cheaply computed, and is shown in Figure~\ref{fig:plummer_likelihood}. Suppose, however, that the likelihood were not cheaply computed. In this case we would proceed as follows.

First we choose the region of parameter space on which we wish to emulate. By the virial theorem we know that $3 \langle v_z^2 \rangle = G M_\mathrm{virial} / r_\textrm{g}$ where $\langle v_z^2 \rangle$ is the line-of-sight velocity dispersion, $M_\mathrm{virial}$ is the virial mass, and $r_\mathrm{g}$ is the gravitational radius \citep{binney_galactic_2008}. We may approximate the gravitational radius by $r_{1 / 2} / 0.45$ \citep{binney_galactic_2008}, where $r_{1 / 2}$ is the half-light radius, and note that for the Plummer sphere, $r_{1 / 2} = b / \sqrt{2^{2 / 3} - 1}$. We might guess that the true value of $M$ is within a factor of three either side of $M_\mathrm{virial}$. Similarly, we might guess that the true value of $b$ is a factor of three either side of its estimate, and that the true value of $r_\mathrm{a}$ is within an order of magnitude either side of its estimate.

However, in the case that the observed data have no error, the feasible region is bounded below by the curve $Q_\mathrm{max} = 0$ (the minimum value $Q$ can take). For a given projected radius, the maximum line-of-sight velocity is therefore set by the condition
\begin{equation}
    (v_z)^2 = \frac{2 \Psi(r) (r_\mathrm{a}^2 + r^2 - r_\mathrm{p}^2)}{r_\mathrm{a}^2 + r^2}
    \label{eq:terminal_velocity}
\end{equation}
where $r_\mathrm{p} \leq r$. We must maximize this expression. The maximum value, $(v_z)_\mathrm{max}$, occurs at a radius determined by the equation 
\begin{align}
    \frac{\dif v_z}{\dif r}\bigg|_{r = r_\mathrm{max}} = 0
    \label{eq:los_vel_max}
\end{align}
which, if it exists, is unique, or (if this equation has no solution on account of the constraint $r \geq r_\mathrm{p}$) at a radius $r = r_\mathrm{p}$.

In the isotropic limit, $r_\mathrm{a} = \infty$, we have $v_z = (v_z)_\mathrm{max}$ at $r = r_\mathrm{p}$, and therefore
\begin{equation}
    \frac{M^2}{(v_z^2 r_\mathrm{p} / 2G)^2} - \frac{b^2}{r_\mathrm{p}^2} = 1,
\end{equation}
a hyperbola in $M$ and $b$. Each pair $(r_\mathrm{p}, v_z)$ defines such a hyperbola. In the anisotropic case, equation~\ref{eq:los_vel_max} gives
\begin{equation}
    r_\mathrm{max}^2 = \frac{(3 r_\mathrm{p}^2 - 2 a^2) - r_\mathrm{p} \sqrt{9 r_\mathrm{p}^2 - 8 r_\mathrm{a}^2 + 8 b^2}}{2}
    \label{eq:radius_max}
\end{equation}
if the discriminant and numerator are real and nonnegative, i.e.\ if
\begin{align}
    r_\mathrm{p} &\ge \frac{r_\mathrm{a}^2}{\sqrt{r_\mathrm{a}^2 + 2b^2}}, \text{ and}\\
    r_\mathrm{a} &\ge b.
\end{align}
Otherwise, $r_\mathrm{max} = r_\mathrm{p}$. In the point-mass limit, $b =0$, and upon substituting equation~\ref{eq:terminal_velocity} into  \ref{eq:los_vel_max} we find an equation in $M$ and $r_\mathrm{a}$. Again, each pair $(r_\mathrm{p}, v_z)$ defines such an equation. A given parameter vector is forbidden if the observed line-of-sight velocity of any star is greater than this maximum allowed velocity.

For our data we find that $\langle v_z^2 \rangle = 504$~km$^2$~s$^{- 2}$, and
$r_{1 / 2} = 0.944$~kpc. Thus $b = 0.723$~kpc and $M_\mathrm{virial} = 0.737
\times 10^9$~M$_{\sun}$. The total mass, $M$, is bounded below by the maximum
value of $v_z^2 r_\mathrm{p} / 2 G$, namely $0.461 \times 10^9$~M$_{\sun}$. We
therefore choose to emulate on the region of parameter space $\bm{A} = [0.461,
  2.21] \times [0.241, 2.17] \times [0.241, 7.23]$. We make a logarithmic
transformation of the parameter space (according to prescription given in
section~\ref{sec:conditioning}), mapping a parameter vector $\bm{a} = (M, b,
r_\mathrm{a})$ to $\bm{x} = (\log M, \log b, \log r_\mathrm{a})$. The
transformed parameter space is $\bm{X} = [-0.336, 0.345] \times [-0.618,
  0.337] \times [-0.618, 0.860]$. We also transform the likelihood (again
according to prescription given in section~\ref{sec:conditioning}) from
$L(\bm{a})$ to $\ln(L_{\bm{X}}(\bm{x}) + \epsilon)$ where $\epsilon = \min
(L_{\bm{X}}(\bm{x}_i))_{i = 1}^N$. We make $10D = 30$ samples from this
transformed parameter space (according the prescription given in
section~\ref{sec:training}) giving the training data $(\bm{x}_i,
\ln(L_{\bm{X}}(\bm{x}_i) + \epsilon))_{i = 1}^N$. We then optimize the model
hyperparameter vector, $\bm{\theta} = (\sigma_\mathrm{SE}^2, m_{\log M},
m_{\log b}, m_{\log{r_\mathrm{a}}})$, using the maximum LOO-likelihood method
(described in section~\ref{sec:validating_the_emulator}), finding that
$\bm{\theta}_* = (535\thinspace000, 23.5, 3.30, 3.43)$, i.e.\ that the length
scales are $l_{\log M} = 0.206$, $l_{\log b} = 0.550$, and
$l_{\log{r_\mathrm{a}}} = 0.540$. We find that the LOOCV score is $R = 0.801$,
and that the extreme value of the LOOCV residuals is 2.33. Diagnostic plots
(Figure~\ref{fig:plummer_function_diagnostics}) show that the standardized
square errors are distributed normally and show no trend across the parameter
space. The results of the validation are acceptable, meaning that we may
proceed to maximize the transformed likelihood using EGO. Using a stopping
threshold of $\epsilon = 0.001$, the EGO algorithm requires $33$ iterations to
find the maximum at $\log M = 0.0356$, $\log b = -0.0288$, and $\log
r_\mathrm{a} = 0.328$. At the last iteration the maximum LOO-likelihood
estimate of the hyperparameter vector is $\bm{\theta} = (203\thinspace000,
17.0, 30.8, 8.10)$, i.e.\ that the length scales are $l_{\log M} = 0.242$,
$l_{\log b} = 0.180$, and $l_{\log{r_\mathrm{a}}} = 0.351$.

We then polish this result by resampling the likelihood in its neighbourhood, and again performing GPE. We choose the region that is within one quarter of a length scale in each parameter, namely $\bm{X}' = [-0.0249, 0.0961] \times [-0.0738, 0.0162] \times [0.240, 0.416]$. We again transform our sample of the likelihood, finding that the most-appropriate transformation is to $\ln L(\bm{x})$, where no offset is required as the likelihood is everywhere defined in this new region of parameter space. The maximum LOO-likelihood estimate of the parameter vector is $\bm{\theta}_* = (436, 24.1, 28.3, 1.42)$, i.e.\ the estimate for the length scales are $l_{\log M} = 0.204$, $l_{\log b} = 0.188$, and $l_{\log{r_\mathrm{a}}} = 0.839$. We find the maximum at $\log M = 0.0327$, $\log b = -0.0213$, and $\log r_\mathrm{a} = 0.305$. We note that for this three-dimensional model we have recovered the MLE with fewer than $100$ evaluations of the likelihood. The first and last sets of 30 evaluations may be each be made in parallel, effectively reducing this number to approximately $40$. Batch-sequential EGO (section~\ref{sec:improving_the_emulator}), would reduce the effective number of runs still further.

The total sensitivity is the initial step of emulation is $\tau = 21.4$, indicating that this problem is hard (section~\ref{sec:training}). We can see why this is the case by inspecting the plot of the likelihood (Figure~\ref{fig:plummer_likelihood}). We note that the likelihood is very sharply peaked. Another way of putting this is to say that it has multiple length scales (the function is highly sensitive to changes in the parameter vector around its maximum, but insensitive to such changes away from its maximum). The squared-exponential covariance function, which assumes a single set of length scales is thus grossly misspecified. Indeed the average separation of the design points is smaller than the peak. The sharpness of this peak is due to several factors: (1) that our data are drawn from the same model we are fitting, (2) that the dimension of our parameter space is small, and (3) that there is no error associated with our synthetic observations. The dynamical model is well-specified and its parameters tightly constrained by the data. The problem of multiple lengths persists even in the transformed data, to which we see an approximation in Figure~\ref{fig:plummer_log_likelihood}. In this case there is a sharp cliff on the boundary of the permitted and forbidden regions of parameter space. Such forbidden regions exist only for data with zero errors. We thus expect the task of fitting this perfectly specified low-dimensional toy model to perfect data to be the maximally difficult case for emulation. We expect it to be considerably harder to than the task of fitting more-sophisticated models to imperfect data, the likelihoods of which will be less sharply peaked, and for which forbidden regions of parameter space do not exist.


\subsection{Confidence region}

\begin{figure}
    \centering
    \includegraphics[width=1.0\linewidth]{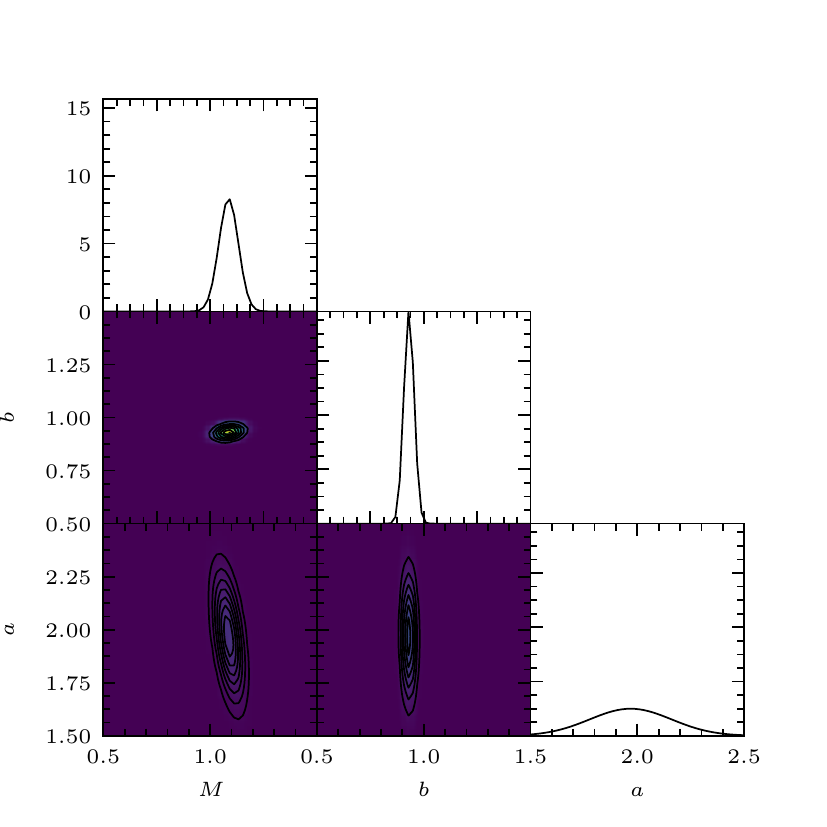}
    \caption{The one- to five-sigma confidence regions for the maximum-likelihood estimate of the Plummer-model parameters, $M$, $b$, and $r_\mathrm{a}$, computed using the Fisher information matrix (equation~\ref{eq:likelihood_confidence}).}
    \label{fig:plummer_param_error}
\end{figure}

\begin{figure}
    \centering
    \includegraphics[width=1.0\linewidth]{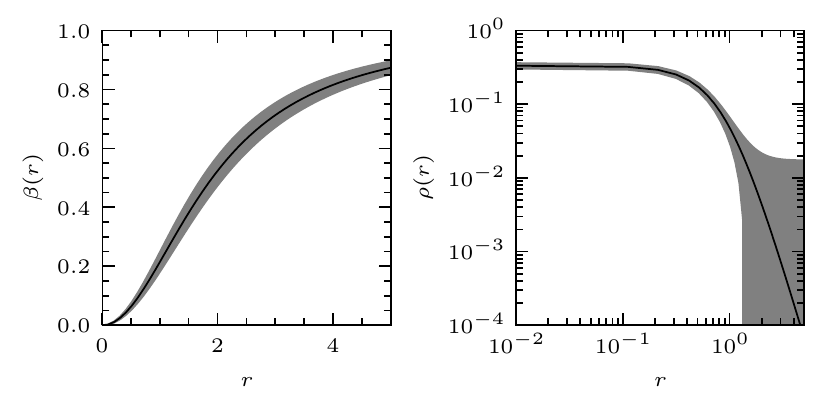}
    \caption{The maximum-likelihood estimates (solid lines) of the Plummer-model density (left) and anisotropy parameter (right), together with their one-sigma confidence intervals (shaded regions).}
    \label{fig:plummer_density_error}
\end{figure}

The Hessian of the log-likelihood is available to us as a consequence of the polishing step (equation~\ref{eq:gpecompute_hessian}). Hence, we may compute an estimate of the Fisher information matrix (equation~\ref{eq:fisher_information}) without further evaluation of the dynamical model. However, our estimator $\hat{r}$ is for the log-likelihood, $\ln L(\bm{x})$, expressed as a function of the transformed parameters, $\bm{x}$. 
Thus, the first derivative is
\begin{align}
    \dfrac{\partial \ln L}{\partial a_j}
    &= \dfrac{\partial \hat{y}}{\partial a_j}\\
    &= \dfrac{\partial \hat{y}}{\partial x_j} \dfrac{\partial x_j}{\partial a_j},
\end{align}
and the second derivative is
\begin{align}
    \dfrac{\partial^2 \ln L}{\partial a_i \partial a_j} 
    &= \dfrac{\partial^2 \hat{y}}{\partial a_i \partial a_j}\\
    &= \dfrac{\partial^2 \hat{y}}{\partial x_i \partial x_j} \dfrac{\partial x_i}{\partial a_i} \dfrac{\partial x_j}{\partial a_j} + \dfrac{\partial \hat{y}}{\partial x_j} \dfrac{\partial^2 x_j}{\partial a_i \partial a_j},
\end{align}
where the second term vanishes at the maximum. Given that $x_i = \log a_i$ we have that
\begin{align}
    \dfrac{\partial x_i}{\partial a_i}
    &= \dfrac{1}{a_i \ln 10}, \text{ and}\\
    \dfrac{\partial^2 x_i}{\partial a_i \partial a_j}
    &= - \dfrac{1}{a_i^2 \ln 10} \delta_{ij}.
\end{align}
In Figure~\ref{fig:plummer_param_error} we plot the confidence regions for the maximum-likelihood estimate of the parameters. The galactic mass, $M$, and scale length, $b$, are well constrained but the anistropy parameter, $r_\mathrm{a}$, is less so. This is as we would expect. For a self-consistent model of this kind the mass and extent of a galaxy are functions of one another through Poisson's equation. All of the observational data therefore contain information about the $M$ and $b$. In the anisotropic case, however, there is an additional length scale, $r_\mathrm{a}$, which we must determine using data at radii greater than this value. Stars at smaller radii do not constrain the length scale, meaning that only a subset of our data contain information about it.

Given the distribution of the MLE for the parameters we may also compute the distribution of the MLE for the density and for Binney's anisotropy parameter using equation \ref{eq:delta_method}. The density is given by equation~\ref{eq:density} and hence the MLE for the density,
\begin{equation}
  \hat{P} \sim N(\rho(\hat{\bm{a}}), \sigma_{\rho}^2),
  \label{eq:density_distribution}
\end{equation}
where
\begin{equation}
    \sigma_{\rho}^2 = \left( \dfrac{\upartial \rho(\hat{\bm{a}})}{\upartial
    \bm{a}}\right)^{\mathrm{t}} \bm{I}_A^{- 1}(\hat{\bm{a}}) \dfrac{\upartial
    \rho(\hat{\bm{a}})}{\upartial \bm{a}}.
    \label{eq:density_distribution}
\end{equation}
For an Ossipkov-Merritt model, Binney's anisotropy parameter \citep{binney_galactic_2008},
\begin{align}
    \beta(r) = \dfrac{1}{1 + r_\mathrm{a}^2 / r^2}.
\end{align}
Hence, the MLE for Binney's anisotropy parameter,
\begin{align}
  \hat{B} &\sim N(\beta(\hat{\bm{a}}), \sigma_{\beta}^2),
  \label{eq:beta_distribution}
\end{align}
where
\begin{equation}
    \sigma_{\beta}^2 = \left( \dfrac{\upartial \beta(\hat{\bm{a}})}{\upartial \bm{a}}
    \right)^{\mathrm{t}} \bm{I}_A^{- 1}(\hat{\bm{a}}) \dfrac{\upartial
    \beta(\hat{\bm{a}})}{\upartial \bm{a}}.
    \label{eq:beta_error}
\end{equation}
We plot the distributions of these quantities in Figure~\ref{fig:plummer_density_error}. These are the pricipal results of our work.

\section{CONCLUSION}
\label{sec:conclusion}

We have presented a novel statistical algorithm for the efficient dynamical modelling of a stellar system. Throughout, our interest has been in what an observational data set can tell us about the stellar system from which it is drawn. In particular, for a dSph, we would like to know which dark-matter morphologies a kinematic data set rules out, and which dark-matter morphologies best account for the data. 

We have adopted the maximum-likelihood approach, which allows us to draw robust confidence intervals within the parameter space of our dynamical model (section 2). The method of maximum likelihood requires us to know the maximum of the likelihood function, and the second derivative of this likelihood function at its maximum, in order that we may compute the Fisher information matrix. Typically, this likelihood function has no closed-form expression and is expensive to evaluate. We have therefore used GPE (\citealp{ohagan_curve_1978} and \citealp{sacks_design_1989}) and efficient global optimization \citep{jones_efficient_1998}, which allow us to optimize the likelihood at significantly reduced computational expense, and to compute good approximations to the second derivatives (section~\ref{sec:gpe}). 

The methods of GPE and EGO are well-established, but there are particular issues in applying them to the situation we have described. The likelihood function is difficult to emulate as it may be sharply peaked. This can cause the GPE to underperform. The solution to this problem is to transform the data so that they better fit the assumptions of the method. This amounts to a nonlinear scaling and reparameterization of the dynamical model followed by validation of the results. Each stage of the analysis may be automated and is implemented in a Python module called PyMimic, which we make publicly available (\href{https://github.com/AmeryGration/pymimic}{https://github.com/AmeryGration/pymimic}). We have given an example of the analysis for the case of a toy model, namely the single-component anisotropic Plummer sphere, which is (counterintuitively) the maximally difficult case. We note this example requires fewer than 100 runs of the dynamical model, and that because the method is trivially parallelizable, the effective number of runs is approximately 40 (if we were to use a naive lattice-based search of the parameter space we might expect to need more than 1000 runs). The method is readily applicable to more-sophisiticated models, and we anticipate that it will allow us to fit a broader class of models than has been computationally tractable. In future work we plan to use it to fit two-component general-profile equilibrium models to observations of the classical dSphs, as well as to fit $N$-body models to observations of tidally-disturbed dSphs \citep{ural_low_2015}.

\section*{Acknowledgements}

We would like to thank Carlos Frenk and Richard Bower for valuable and insightful discussions. Their work prompted this research. We would also like to thank Sylvy Anscombe for numerous discussions of the mathematical structure of GPE.

This work used the DiRAC Complexity system, operated by the University of Leicester IT Services, which forms part of the STFC DiRAC HPC Facility (www.dirac.ac.uk ). This equipment is funded by BIS National E-Infrastructure capital grant ST/K000373/1 and  STFC DiRAC Operations grant ST/K0003259/1. DiRAC is part of the National E-Infrastructure.

\bibliographystyle{mnras}
\bibliography{bibliography}

\begin{thebibliography}{}
\makeatletter
\relax
\def\mn@urlcharsother{\let\do\@makeother \do\$\do\&\do\#\do\^\do\_\do\%\do\~}
\def\mn@doi{\begingroup\mn@urlcharsother \@ifnextchar [ {\mn@doi@}
  {\mn@doi@[]}}
\def\mn@doi@[#1]#2{\def\@tempa{#1}\ifx\@tempa\@empty \href
  {http://dx.doi.org/#2} {doi:#2}\else \href {http://dx.doi.org/#2} {#1}\fi
  \endgroup}
\def\mn@eprint#1#2{\mn@eprint@#1:#2::\@nil}
\def\mn@eprint@arXiv#1{\href {http://arxiv.org/abs/#1} {{\tt arXiv:#1}}}
\def\mn@eprint@dblp#1{\href {http://dblp.uni-trier.de/rec/bibtex/#1.xml}
  {dblp:#1}}
\def\mn@eprint@#1:#2:#3:#4\@nil{\def\@tempa {#1}\def\@tempb {#2}\def\@tempc
  {#3}\ifx \@tempc \@empty \let \@tempc \@tempb \let \@tempb \@tempa \fi \ifx
  \@tempb \@empty \def\@tempb {arXiv}\fi \@ifundefined
  {mn@eprint@\@tempb}{\@tempb:\@tempc}{\expandafter \expandafter \csname
  mn@eprint@\@tempb\endcsname \expandafter{\@tempc}}}

\bibitem[\protect\citeauthoryear{Adler}{Adler}{2010}]{adler_geometry_2010}
Adler R.~J.,  2010, The {Geometry} of {Random} {Fields}.
Society for Industrial and Applied Mathematics, Philadelphia, Pa

\bibitem[\protect\citeauthoryear{Bachoc}{Bachoc}{2013}]{bachoc_cross_2013}
Bachoc F.,  2013

\bibitem[\protect\citeauthoryear{Battaglia, Helmi, Tolstoy, Irwin, Hill  \&
  Jablonka}{Battaglia et~al.}{2008}]{battaglia_kinematic_2008}
Battaglia G.,  Helmi A.,  Tolstoy E.,  Irwin M.,  Hill V.,   Jablonka P.,
  2008, The Astrophysical Journal, 681, L13

\bibitem[\protect\citeauthoryear{Binney \& Tremaine}{Binney \&
  Tremaine}{2008}]{binney_galactic_2008}
Binney J.,  Tremaine S.,  2008, Galactic {Dynamics}: {Second} {Edition}, second
  edition edn.
Princeton University Press, Princeton

\bibitem[\protect\citeauthoryear{Bower, Vernon, Goldstein, Benson, Lacey,
  Baugh, Cole  \& Frenk}{Bower et~al.}{2010}]{bower_parameter_2010}
Bower R.~G.,  Vernon I.,  Goldstein M.,  Benson A.~J.,  Lacey C.~G.,  Baugh
  C.~M.,  Cole S.,   Frenk C.~S.,  2010, Monthly Notices of the Royal
  Astronomical Society, 407, 2017

\bibitem[\protect\citeauthoryear{Box \& Cox}{Box \&
  Cox}{1964}]{box_analysis_1964}
Box G. E.~P.,  Cox D.~R.,  1964, Journal of the Royal Statistical Society.
  Series B (Methodological), 26, 211

\bibitem[\protect\citeauthoryear{Breddels \& Helmi}{Breddels \&
  Helmi}{2013}]{breddels_model_2013}
Breddels M.~A.,  Helmi A.,  2013, 558, A35

\bibitem[\protect\citeauthoryear{Carollo, de Zeeuw  \& van~der Marel}{Carollo
  et~al.}{1995}]{carollo_velocity_1995}
Carollo C.~M.,  de Zeeuw P.~T.,   van~der Marel R.~P.,  1995, Monthly Notices
  of the Royal Astronomical Society, 276

\bibitem[\protect\citeauthoryear{Draguljić, Santner  \& Dean}{Draguljić
  et~al.}{2012}]{draguljic_noncollapsing_2012}
Draguljić D.,  Santner T.~J.,   Dean A.~M.,  2012, Technometrics, 54, 169

\bibitem[\protect\citeauthoryear{Draper \& Cox}{Draper \&
  Cox}{1969}]{draper_distributions_1969}
Draper N.~R.,  Cox D.~R.,  1969, Journal of the Royal Statistical Society.
  Series B (Methodological), 31, 472

\bibitem[\protect\citeauthoryear{Evans, Aigrain, Gibson, Barstow, Amundsen,
  Tremblin  \& Mourier}{Evans et~al.}{2015}]{evans_uniform_2015}
Evans T.~M.,  Aigrain S.,  Gibson N.,  Barstow J.~K.,  Amundsen D.~S.,
  Tremblin P.,   Mourier P.,  2015, Monthly Notices of the Royal Astronomical
  Society, 451, 680

\bibitem[\protect\citeauthoryear{Gibson, Aigrain, Roberts, Evans, Osborne  \&
  Pont}{Gibson et~al.}{2012}]{gibson_gaussian_2012}
Gibson N.~P.,  Aigrain S.,  Roberts S.,  Evans T.~M.,  Osborne M.,   Pont F.,
  2012, Monthly Notices of the Royal Astronomical Society, 419, 2683

\bibitem[\protect\citeauthoryear{Heitmann, Higdon, White, Habib, Williams  \&
  Wagner}{Heitmann et~al.}{2009}]{heitmann_coyote_2009}
Heitmann K.,  Higdon D.,  White M.,  Habib S.,  Williams B.~J.,   Wagner C.,
  2009, The Astrophysical Journal, 705, 156

\bibitem[\protect\citeauthoryear{Jones, Schonlau  \& Welch}{Jones
  et~al.}{1998}]{jones_efficient_1998}
Jones D.~R.,  Schonlau M.,   Welch W.~J.,  1998, Journal of Global
  Optimization, 13, 455

\bibitem[\protect\citeauthoryear{Loeppky, Sacks  \& Welch}{Loeppky
  et~al.}{2009}]{loeppky_choosing_2009}
Loeppky J.~L.,  Sacks J.,   Welch W.~J.,  2009, Technometrics, 51, 366

\bibitem[\protect\citeauthoryear{{Lovell}, {Frenk}, {Eke}, {Jenkins}, {Gao}  \&
  {Theuns}}{{Lovell} et~al.}{2014}]{lovell_2014}
{Lovell} M.~R.,  {Frenk} C.~S.,  {Eke} V.~R.,  {Jenkins} A.,  {Gao} L.,
  {Theuns} T.,  2014, \mnras, 439, 300

\bibitem[\protect\citeauthoryear{{Ludlow}, {Bose}, {Angulo}, {Wang},
  {Hellwing}, {Navarro}, {Cole}  \& {Frenk}}{{Ludlow}
  et~al.}{2016}]{ludlow_2016}
{Ludlow} A.~D.,  {Bose} S.,  {Angulo} R.~E.,  {Wang} L.,  {Hellwing} W.~A.,
  {Navarro} J.~F.,  {Cole} S.,   {Frenk} C.~S.,  2016, \mnras, 460, 1214

\bibitem[\protect\citeauthoryear{Mashchenko, Wadsley  \& Couchman}{Mashchenko
  et~al.}{2008}]{mashchenko_stellar_2008}
Mashchenko S.,  Wadsley J.,   Couchman H. M.~P.,  2008, Science (New York,
  N.Y.), 319, 174

\bibitem[\protect\citeauthoryear{McKay, Beckman  \& Conover}{McKay
  et~al.}{1979}]{mckay_comparison_1979}
McKay M.~D.,  Beckman R.~J.,   Conover W.~J.,  1979, Technometrics, 21, 239

\bibitem[\protect\citeauthoryear{Merritt}{Merritt}{1985}]{merritt_spherical_1985}
Merritt D.,  1985, The Astronomical Journal, 90, 1027

\bibitem[\protect\citeauthoryear{Moore, Berry, Chua  \& Gair}{Moore
  et~al.}{2016}]{moore_improving_2016}
Moore C.~J.,  Berry C. P.~L.,  Chua A. J.~K.,   Gair J.~R.,  2016, Physical
  Review D, 93

\bibitem[\protect\citeauthoryear{Navarro, Eke  \& Frenk}{Navarro
  et~al.}{1996}]{navarro_cores_1996}
Navarro J.~F.,  Eke V.~R.,   Frenk C.~S.,  1996, Monthly Notices of the Royal
  Astronomical Society, 283, L72

\bibitem[\protect\citeauthoryear{O'Hagan \& Kingman}{O'Hagan \&
  Kingman}{1978}]{ohagan_curve_1978}
O'Hagan A.,  Kingman J. F.~C.,  1978, Journal of the Royal Statistical Society.
  Series B (Methodological), 40, 1

\bibitem[\protect\citeauthoryear{Osipkov}{Osipkov}{1979}]{osipkov_spherical_1979}
Osipkov L.~P.,  1979, Pisma v Astronomicheskii Zhurnal, 5, 77

\bibitem[\protect\citeauthoryear{Plummer}{Plummer}{1911}]{plummer_problem_1911}
Plummer H.~C.,  1911, Monthly Notices of the Royal Astronomical Society, 71,
  460

\bibitem[\protect\citeauthoryear{Rasmussen \& Williams}{Rasmussen \&
  Williams}{2006}]{rasmussen_gaussian_2006}
Rasmussen C.~E.,  Williams C. K.~I.,  2006, Gaussian processes for machine
  learning.
MIT Press, Cambridge, Mass

\bibitem[\protect\citeauthoryear{Read \& Gilmore}{Read \&
  Gilmore}{2005}]{read_mass_2005}
Read J.~I.,  Gilmore G.,  2005, Monthly Notices of the Royal Astronomical
  Society, 356, 107

\bibitem[\protect\citeauthoryear{Read \& Steger}{Read \&
  Steger}{2017}]{read_how_2017}
Read J.~I.,  Steger P.,  2017, Monthly Notices of the Royal Astronomical
  Society, 471, 4541

\bibitem[\protect\citeauthoryear{Read, Walker  \& Steger}{Read
  et~al.}{2018}]{read_case_2018}
Read J.~I.,  Walker M.~G.,   Steger P.,  2018, arXiv:1805.06934 [astro-ph]

\bibitem[\protect\citeauthoryear{Sacks, Welch, Mitchell  \& Wynn}{Sacks
  et~al.}{1989}]{sacks_design_1989}
Sacks J.,  Welch W.~J.,  Mitchell T.~J.,   Wynn H.~P.,  1989, Statistical
  Science, 4, 409

\bibitem[\protect\citeauthoryear{Sale \& Magorrian}{Sale \&
  Magorrian}{2014}]{sale_three-dimensional_2014}
Sale S.~E.,  Magorrian J.,  2014, Monthly Notices of the Royal Astronomical
  Society, 445, 256

\bibitem[\protect\citeauthoryear{Sampson \& Guttorp}{Sampson \&
  Guttorp}{1992}]{sampson_nonparametric_1992}
Sampson P.~D.,  Guttorp P.,  1992, Journal of the American Statistical
  Association, 87, 108

\bibitem[\protect\citeauthoryear{Santner, Williams  \& Notz}{Santner
  et~al.}{2003}]{santner_design_2003}
Santner T.~J.,  Williams B.~J.,   Notz W.~I.,  2003, The {Design} and
  {Analysis} of {Computer} {Experiments}, 2003 edition edn.
Springer, New York

\bibitem[\protect\citeauthoryear{Schonlau \& Welch}{Schonlau \&
  Welch}{1996}]{schonlau_global_1996}
Schonlau M.,  Welch W.~J.,  1996, in , Proceedings of the {ASA}.
American Statistical Association, pp 183--186

\bibitem[\protect\citeauthoryear{Strigari, Bullock, Kaplinghat, Simon, Geha,
  Willman  \& Walker}{Strigari et~al.}{2008}]{strigari_common_2008}
Strigari L.~E.,  Bullock J.~S.,  Kaplinghat M.,  Simon J.~D.,  Geha M.,
  Willman B.,   Walker M.~G.,  2008, Nature, 454, 1096

\bibitem[\protect\citeauthoryear{Strigari, Frenk  \& White}{Strigari
  et~al.}{2010}]{strigari_kinematics_2010}
Strigari L.~E.,  Frenk C.~S.,   White S. D.~M.,  2010, Monthly Notices of the
  Royal Astronomical Society, 408, 2364

\bibitem[\protect\citeauthoryear{Sundararajan \& Keerthi}{Sundararajan \&
  Keerthi}{2001}]{sundararajan_predictive_2001}
Sundararajan S.,  Keerthi S.~S.,  2001, Neural Computation, 13, 1103

\bibitem[\protect\citeauthoryear{Ural, Wilkinson, Read  \& Walker}{Ural
  et~al.}{2015}]{ural_low_2015}
Ural U.,  Wilkinson M.~I.,  Read J.~I.,   Walker M.~G.,  2015, Nature
  Communications, 6

\bibitem[\protect\citeauthoryear{Vazquez \& Bect}{Vazquez \&
  Bect}{2010}]{vazquez_convergence_2010}
Vazquez E.,  Bect J.,  2010, Journal of Statistical Planning and Inference,
  140, 3088

\bibitem[\protect\citeauthoryear{Walker, Mateo, Olszewski, Peñarrubia, Evans
  \& Gilmore}{Walker et~al.}{2009}]{walker_universal_2009}
Walker M.~G.,  Mateo M.,  Olszewski E.~W.,  Peñarrubia J.,  Evans N.~W.,
  Gilmore G.,  2009, The Astrophysical Journal, 704, 1274

\bibitem[\protect\citeauthoryear{Wasserman}{Wasserman}{2007}]{wasserman_all_2007}
Wasserman L.,  2007, All of {Nonparametric} {Statistics}: {A} {Concise}
  {Course} in {Nonparametric} {Statistical} {Inference}.
Springer, New York

\bibitem[\protect\citeauthoryear{Wilkinson, Kleyna, Evans  \&
  Gilmore}{Wilkinson et~al.}{2002}]{wilkinson_dark_2002}
Wilkinson M.~I.,  Kleyna J.,  Evans N.~W.,   Gilmore G.,  2002, Monthly Notices
  of the Royal Astronomical Society, 330, 778

\bibitem[\protect\citeauthoryear{Wolf, Martinez, Bullock, Kaplinghat, Geha,
  Munoz, Simon  \& Avedo}{Wolf et~al.}{2010}]{wolf_accurate_2010}
Wolf J.,  Martinez G.~D.,  Bullock J.~S.,  Kaplinghat M.,  Geha M.,  Munoz
  R.~R.,  Simon J.~D.,   Avedo F.~F.,  2010, Monthly Notices of the Royal
  Astronomical Society

\makeatother
\end{thebibliography}

\appendix

\section{Derivatives of Gaussian processes}

\label{appendix1}

For convenience we introduce the vector $\bm{\alpha} := \bm {K}^{-1}\bm {y}$. From equation~\ref{eq:gpe_mean} we find that the gradient of the mean,
\begin{align}
    \dfrac{\partial \hat{y}(\bm {a})}{\partial \bm {a}} = \dfrac{\partial \bm {k}(\bm {a})}{\partial \bm {a}} \bm{\alpha},
    \label{eq:gradient}
\end{align}
or, in component form,
\begin{align}
    \dfrac{\partial \hat{y}(\bm{a})}{\partial a_j} = \sum_{i = 1}^{N} \dfrac{\partial k_i(\bm{a})}{\partial a_j} \alpha_{i}.
\end{align}
The Hessian of the mean is then 
\begin{align}
    \dfrac{\partial^2 \hat{y}(\bm {a})}{\partial \bm {a} \partial \bm{a}^\mathrm{t}} = \dfrac{\partial^2 \bm {k}(\bm {a})}{\partial \bm{a} \partial \bm{a}^\mathrm{t}} \bm{\alpha},
    \label{eq:hessian}
\end{align}
or, in component form,
\begin{align}
    \dfrac{\partial^2 \hat{y}(\bm{a})}{\partial a_i \partial a_j} = \sum_{i = 1}^N \dfrac{\partial^2 k_i(\bm{a})}{\partial a_i \partial a_j} \alpha_i.
\end{align}

\subsection{Squared-exponential kernel}
\label{sec:se_derives}

The squared-exponential kernel, 
\begin{align}
    k_\mathrm{SE}(\bm {a}, \bm {a}') := \sigma_\mathrm{SE}^2 \exp \left( -\dfrac{1}{2} (\bm {a} - \bm {a}')^\mathrm{t} \bm {M} (\bm {a} - \bm {a}') \right).
\end{align}
Hence
\begin{align}
    \dfrac{\partial k_\mathrm{SE}(\bm {a}, \bm {a}')}{\partial \bm {a}}
    &= - k_\mathrm{SE}(\bm {a}, \bm {a}') \bm {M} (\bm {a} - \bm {a}'),\\
    \dfrac{\partial k_\mathrm{SE}(\bm {a}, \bm {a}')}{\partial \bm {a}'}
    &= - \dfrac{\partial k_\mathrm{SE}(\bm {a}, \bm {a}')}{\partial \bm{a}},\\
    \dfrac{\partial^2 k_\mathrm{SE}(\bm {a}, \bm {a}')}{\partial \bm{a} \partial \bm{a}^\mathrm{t}}
    &= k_\mathrm{SE}(\bm {a}, \bm {a}') \bm {M} \left((\bm{a} - \bm {a}') (\bm {a} - \bm {a}')^\mathrm{t} \bm {M} - \bm {I}\right),\\
    \dfrac{\partial^2 k_\mathrm{SE}(\bm {a}, \bm {a}')}{\partial \bm{a}' \partial {\bm{a}'}^\mathrm{t}}
    &=     \dfrac{\partial^2 k_\mathrm{SE}(\bm {a}, \bm {a}')}{\partial \bm {a} \partial \bm{a}^\mathrm{t}},
\end{align}
and
\begin{align}
    \dfrac{\partial^2 k_\mathrm{SE}(\bm {a}, \bm {a}')}{\partial \bm{a} \partial {\bm{a}'}^\mathrm{t}} &= - \dfrac{\partial^2 k_\mathrm{SE}(\bm {a}, \bm {a}')}{\partial \bm{a} \partial \bm{a}^\mathrm{t}}.
\end{align}
Thus we may compute the derivatives given in equations~\ref{eq:gradient} and \ref{eq:hessian}.



\bsp	
\label{lastpage}

\end{document}